\pdfoutput=1
\documentclass[10pt,preprint]{aastex}

\shorttitle{Stellar Populations of Four Distant Compact Galaxies.}
\shortauthors{Hoyos et al.}

\begin{document}

\title{Stellar Populations Found in the Central kpc of Four Luminous Compact Blue Galaxies at
  Intermediate \textit{z}.\footnote{Based on observations obtained with
  the NASA/ESA {\em Hubble Space Telescope} through the Space Telescope
Science Institute which is operated by Association of Universities for 
Research in Astronomy (AURA), Inc., under NASA contract NAS 5-26555}}

\author{C. Hoyos.\altaffilmark{1}\email{charly.hoyos@uam.es} R. Guzm\'an \altaffilmark{2}\email{guzman@astro.ufl.edu} A. I. D\'{\i}az\altaffilmark{1,5}\email{angeles.diaz@uam.es} D. C. Koo\altaffilmark{3}\email{koo@ucolick.org} M. A. Bershady.\altaffilmark{4}\email{mab@astro.wisc.edu} }

\altaffiltext{1}{Departamento de F\'{\i}sica Te\'orica (C-XI), Universidad Aut\'onoma de Madrid. Carretera de Colmenar Viejo km 15.600 28049 Madrid, Spain.}
\altaffiltext{2}{Bryant Space Science Center, University of Florida, Gainesville, FL 32611-2055}
\altaffiltext{3}{Department of Astronomy, University of California Santa Cruz, Santa Cruz, CA 95064}
\altaffiltext{4}{Department of Astronomy, University of Wisconsin, Madison, WI 53706}
\altaffiltext{5}{On sabbatical leave at IoA, Cambridge, UK.}

\begin{abstract}

We investigate the star formation history of the central regions of four Luminous Compact
Blue Galaxies (LCBGs) at intermediate redshift using evolutionary population synthesis techniques.
LCBGs are blue ($\bv \leq 0.6$), compact ($\mu_{B}\leq 21.0$ mag arcsec$^{-2}$) galaxies with 
absolute magnitudes M$_{B}$ brighter than -17.5.
The LCBGs analyzed here are located at $0.436 \leq z \leq 0.525$.
They are among the most luminous (M$_{B} < -20.5$), blue ($\bv \leq 0.4$) and high surface 
brightness ($\mu_{B}\leq 19.0$ mag arcsec$^{-2}$) of this population.
The observational data used were obtained with the HST/STIS spectrograph, the HST/WF/PC-2 camera 
and the HST/NICMOS first camera.
We have disentangled the stellar generations found in the central regions of the 
observed targets using a very simple model. This is one of the first times this 
is done for compact galaxies at this redshift using Hubble Space Telescope data, and 
it provides a comparison bench for the future work on this kind of galaxies using 
instruments with adaptive optics in 10-m class telescopes.
We find evidence for multiple stellar populations. One of them is identified as 
the ionizing population, and the other one corresponds to the underlying stellar generation.
The estimated masses of the inferred stellar populations are compatible with the dynamical
 masses, which are typically 2--10$\times 10^{9}M_{\odot}$.
Our models also indicate that the first episodes of star formation the presented LCBGs underwent happened
between 5 and 7 Gyr ago.
We compare the stellar populations found in LCBGs with the stellar populations present in 
bright, local \ion{H}{2} galaxies, nearby spheroidal systems and Blue Compact Dwarf 
Galaxies. It turns out that the underlying stellar populations of LCBGs are similar yet 
bluer to those of local H\textsc{II} galaxies. It is also the case that the passive color 
evolution of the LCBGs could convert them into local Spheroidal galaxies if no further 
episode of star formation takes place. Our results help to impose constraints on evolutionary 
scenarios for the population of LCBGs found commonly at intermediate redshifts.

\end{abstract}


\section{Introduction}

Luminous Blue Compact Galaxies (LCBGs) are luminous (M$_{B}$ brighter than -17.5), blue (\bv $\leq$ 0.6), and 
compact ($\mu_{B}\leq 21.0$ mag arcsec$^{-2}$) systems experiencing an intense episode of star formation, with
typical star formation rates ranging from 1 to 5 $\textrm{M}_{\odot}\textrm{yr}^{-1}$. Although these star
formation rates are common in other starburst systems, the star formation episode takes place in a very small
volume, allowing these compact objects to be seen from cosmological distances. LCBGs are believed 
to contribute up to $\sim$50\% of the star formation rate density in the universe
at $z\approx 1$ \citep{guz97}. Their rapid evolution from those times till 
the present day makes them responsible for the observed evolution of the luminosity function \citep{lil95}
as well as the evolution of the star formation rate in the universe.

Despite their importance, their relation to local galaxies is poorly known. The works presented 
in  \citet{guz97,phi97,hoyos04} give a glimpse of the similarities between LCBGs and the local 
\mbox{H\textsc{II}} and Starburst Nuclei populations. \mbox{H\textsc{II}} galaxies are a subset of 
Blue Compact Galaxies in which the spectrum is completely dominated by a component, present 
almost everywhere within the galaxy, resembling the emission of a \mbox{H\textsc{II}} 
region \citep{sargearl70}. Blue Compact Galaxies (BCGs) were first identified by \citet{zwick65} as faint 
star-like field galaxies on Palomar Sky Survey plates. BCGs often present a emission line spectrum 
and a UV excess. Starburst Nuclei were introduced in \citet{balza83}; they show high 
extinction values, with very low [\mbox{N\textsc{II}}]$\lambda$ 6584/H$\alpha$ ratios and faint 
[\mbox{O\textsc{III}}]$\lambda 5007$ emission. Their H$\alpha$ luminosities are always greater 
than $10^{8}$ solar luminosities. One of the main ingredients in any theory attempting to explain
 the final destiny of LCBGs is bound to be an accurate knowledge of
the stellar generations already existing in these distant sources. The similarities found between LCBGs and 
many local \mbox{H\textsc{II}} galaxies make it possible that the star formation histories of
both types of objects are very similar. Even though \ion{H}{2} galaxies were originally 
thought to be truly young objects, experiencing their very first star formation
 episodes \citep{sargearl72,sargearl70}, this hypothesis is no longer tenable after very deep CCD imaging 
of such objects \citep{vri_telles97,etelles97}, or other very similar sources like the
Blue Compact Dwarf Galaxies (BCD). BCDs are similar to BCGs or
H\textsc{II} galaxies, but are an order of magnitude less luminous. The underlying stellar
populations of BCDGs are also older than those of BCGs. (\citealp{papa96,loosethuan86}, or \citealp{papa96_bis}) 
which show clear evidence of an older extended population of stars. The properties of this more extended
component has been further studied photometrically 
\citep[see, e.g.][]{caon05,etelles97}.
The current view on local \ion{H}{2} galaxies
is that they are a mixture of three populations. The first of them is the youngest
population, responsible for the observed emission lines. The second population
is an intermediate age generation, older than 50Myr but younger than 1 Gyr. This population
is unable to ionize the \ion{H}{1} gas, and its luminosity is increasingly dominated by 
the contribution of giant stars, and  by the contribution of some Asymptotic Giant Branch stars.
The third population is the oldest one, older than 1Gyr. This population generally dominates the stellar mass by a
large factor, and it also occupies a larger volume.

The main target of this study
is to characterize the stellar populations present in intermediate redshift LCBGs by combining the
spectral and color information available from these data with evolutionary population synthesis techniques. 
The modelled stellar populations are then compared to the stellar populations found in 
\mbox{H\textsc{II}} galaxies and the smaller yet similar Blue Compact Dwarf galaxies. It is also 
interesting to study the future evolution of the predicted stellar populations, to see whether or
not they can become dwarf elliptical systems (dE, or Sph), as suggested by \citet{koo95,guz98} 
or, alternatively, they can be the bulges of spiral galaxies still in their 
formation process \citep{ham01,koza99,guz98,phi97}.

The observations and analysis techniques together with the different results they yield are described in 
section \ref{ODR}. Section \ref{resultados} presents the data analysis. The models that have 
been constructed to explain the observations are explained in section \ref{calcus}.
The LCBGs presented in this work are compared to relevant samples of local galaxies in section
\ref{compa}. The summary is given in section \ref{summ}.

The cosmology assumed here is a flat universe with $\Omega_{\Lambda}=0.7$ and $\Omega_{matter}=0.3$. The 
resulting cosmological parameters are: H$_{0}$=70 km s$^{-1}$Mpc$^{-1}$, q$_{0}$=-0.55. Given these 
values, $1\arcsec$ corresponds to 1.8kpc at $z=0.1$ and to 5.8 kpc at $z=0.45$.

\section{Observations and Data Reduction.} \label{ODR}

In this paper we present long-slit spectra, deep optical
V (F606W) and I (F814W) images, and infrared F160W images of four LCBGs with redshifts between
0.436 and 0.525. The data were obtained using the STIS instrument, the
WF/PC-2 camera and the NICMOS first camera.
Target names, redshifts, look-back times, \textit{B} absolute
magnitudes, exposure times, half light-radii, velocity dispersions and virial
masses within $R_{e}$, as well as an
identifying numeral for the observed objects can be seen in table
\ref{prestab}.

\begin{deluxetable}{lllll}
\tablecolumns{5}
\tablewidth{0pc}
\tablecaption{Log of HST/STIS, HST/WF/PC-2 and HST/NIC1 
observations.\label{prestab}}

\tablehead{
\colhead{Parameter.} & \colhead{SA57-7042.} & \colhead{SA57-5482}  & \colhead{SA57-10601.}  &  \colhead{H1-13088.} 
}
\startdata

RA\tablenotemark{a}      & 13 07 26.3   & 13 09 08.8   & 13 08 47.8 & 17 20 19.67  \\
DEC\tablenotemark{a}     & 29 18 25.7   & 29 15 57.7   & 29 23 41.1 & 50 01 04.7   \\
Obs Date. 	         & 2002-05-05   & 2002-04-09   & 2000-07-05 & 2000-11-06   \\
G750L Exp. Time.         & 9898  	& \nodata      & 4696       & 4932         \\
G750M Exp. Time.         & 2760  	& 18106        & 5571       & 5947         \\
G430L Exp. Time.         & \nodata      & 2790         & \nodata    & \nodata      \\
G430M Exp. Time.         & \nodata      & 7709         & \nodata    & \nodata      \\
WF/PC-2 F606W Exp. Time. & 700   	& 700          & 700        & 700          \\
WF/PC-2 F814W Exp. Time. & 900   	& 900          & 900        & 1000         \\
NICMOS F160W Exp. Time.  & 2560         & 1280         & 5632       & 1792         \\
Obj ID.    	         & 1  	        & 2            & 3          & 4           \\
\textit{z.} 	         & 0.525        & 0.453        & 0.438      & 0.436      \\
$t_{look-back}$(Gyr)     & 5.2  	& 4.7          & 4.6        & 4.5        \\
M$_{B}$                  &   -20.6      & -20.8        & -20.4      & -21.1      \\  
R$_{e}$\tablenotemark{b} & 1.4	        & 1.5          & 1.9        & 2.4        \\  
m$_{B}$                  & 22.3  	& 21.6         & 21.6       & 21.0       \\
r$_{e}$\tablenotemark{c} & 0.25 	& 0.29         & 0.36       & 0.47       \\
Vel. Disp. (km s$^{-1}$  & 120$\pm$10  	& 60$\pm$6         & 42$\pm$4       & 47$\pm$4     \\
$\log M/M_{\odot}$ \tablenotemark{d}    & 10.7 	& 10.1       & 9.7       & 9.8       \\

\enddata
\tablenotetext{a}{Coordinates given for equinox J2000.}
\tablenotetext{b}{Half-light radii in kpc.}
\tablenotetext{c}{Half-light radii in arcseconds.}
\tablenotetext{d}{Dynamical masses within $R_{e}$ calculated as in \citet{hoyos04}. Uncertainties are 20\%.}
\tablecomments{All quantities given were calculated assuming $q_{0}=-.55$ and $H_{0}=70$km s$^{-1}$Mpc$^{-1}$.}
\end{deluxetable}

The STIS long slit spectra are imaged on a SITe 1024x1024 CCD, with $
0.05 \arcsec$ square pixels operating from $\sim$2000 to 11000
\AA. Several instrumental configurations were used. The instrumental
setup used to observe objects 1, 3 \& 4 used the low resolution
grating G750L.  This instrumental setup provides a dispersion of 4.92
\AA{} per pixel or about 190 km s$^{-1}$ per pixel which, in
combination with a slit $0.5 \arcsec$ wide, gives a spectral
resolution of about 29\AA(FWHM). The spectral coverage provided by
this setup is about 5000 \AA. The central wavelength is 7750 \AA.  The
higher resolution grating G750M was also used for objects 3 \& 4. It
provides a dispersion of 0.53 \AA{} per pixel or about 35 km s$^{-1}$
per pixel. The slit width used was $0.2 \arcsec$ wide, thus giving a
spectral resolution of about 2.1\AA (FWHM). The spectral coverage
provided by this setup is about 500 \AA. These spectra are used to
study the gas-phase velocity field, and the results from this work are
presented in \citet{bershady06}.  Object 2 was observed using the
lower-resolution grating G430M and the high-resolution grating
G750M. The G430M grating has a dispersion of 2.7 \AA{} per pixel and
the slit width used is $0.2\arcsec$. The central wavelength is 4300
\AA {}, and the spectral coverage is 2800 \AA. The data were reduced
using the STSDAS\footnote{(Space Telescope Science Data Analysis
System) from the Space Telescope Science Institute, operated for NASA
by AURA} package within IRAF
\footnote{Image Reduction and Analysis Facility. Distributed by the
  National Optical Astronomy Observatories, which is operated by AURA
  (Association of Universities for Research in Astronomy, Inc) under
  cooperative agreement with the National Science
  Foundation.}. For a complete account of the reduction procedures, uncertainties
 and detailed descriptions of the spectra, see \citet{hoyos04}.
It is worth mentioning that STIS spectroscopic observations using the
G750L and G750M gratings are affected by fringing in the red end of the
spectrum. This makes it neccessary to use specifically designed flats
(called ``contemporaneous flats'') in order to circumvent this issue.
Thes use of this technique, which is implemented as a task within STSDAS
raises the S/N by around 15\%. This is crucial for these observations.
The proposal IDs for these observations are GO8678 and GO9126.

The images used were taken with the WF/PC-2 instrument, with the LCBGs
centered in the Planetary Camera, a $800 \times 800$ CCD, with
0.046$\arcsec$ square pixels. Two broadband filters were used, the
F814W filter and the F606W filter.  For each object and filter, two
$\sim 400$s exposures were obtained, to allow for cosmic ray
rejection. The details of the reduction and analysis procedures can be
found in \citet{guz98}. The proposal ID for these data is GO5994.

The infrared frames were obtained with the NIC1 camera of the 
NICMOS instrument\footnote{Before the cryo cooler was installed.}.
This camera uses a low-noise, high QE, 256$\times$256 pixel HgCdTe array, with
0.0431$\arcsec$ square pixels. The filter used was the F160W, which has
a bandpass of around 4000\AA. For each object, several different exposures were collected 
and combined to produce a final mosaic. The proposal ID is GO7875.

The STIS data can only probe the central regions of these galaxies because the outermost regions
of these sources are too faint to be studied spectroscopically with HST. Specifically, the STIS data can be used
to study the line-emitting region, where the starburst is taking place and most of the gas is ionized.
The line-emitting region of the observed galaxies has a radius between 1.4 and 2.0 kpc. This is 
a little bit less than one effective radius.
On the other hand, the WF/PC-2 images and NICMOS frames are deeper, and the size of the observed galaxies in the images
is much larger than the size of the line emittiong region in the STIS spectra. The analysis presented
in this paper focuses on this particular central region where the STIS spectra spatially overlaps with
the WF/PC-2 images and NICMOS frames. This region has a radius of 1.7 kpc, on average.

\section{Results and Analysis.} 
\label{resultados}

\subsection{Emission-Line and Continuum Distribution.}

It is interesting to investigate whether or not the continuum
distribution and the emission-line distribution differ in the STIS
spectra. This will help determine where the starburst has taken
place. This information can be helpful to understand the starburst
triggering mechanism, and to construct models for these objects. Here,
the analysis presented in \citet{hoyos04} is followed, although it is
presented in a much simpler way. Furthermore, only objects 1 \& 2 are
presented here since the other two sources were dealt with in
\citet{hoyos04}. The procedure is briefly reviewed.

\begin{figure}
\plotone{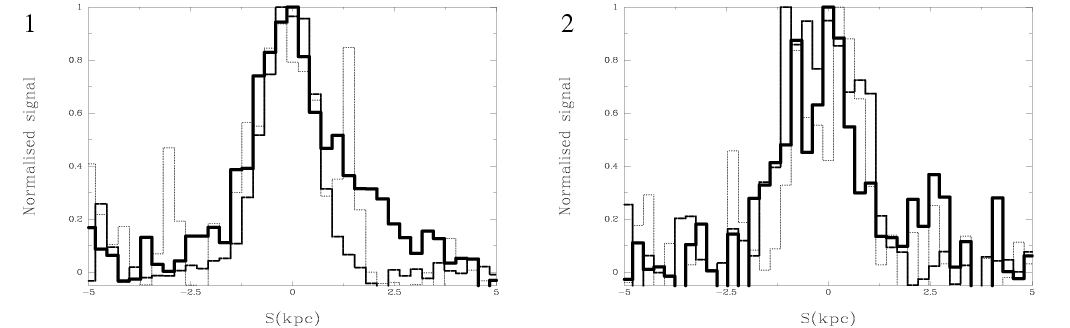}
\caption{Continuum and emission-line spatial profiles. Only objects 1
\& 2 are presented since the spatial profiles for objects 3 \& 4 were
given in \citet{hoyos04}.  The S-axis is placed with respect to the
slits depicted in Fig.\ref{registro}.  Increasing S corresponds to
upper positions in that figure. The continuum is the thickest, solid
line; [\ion{O}{3}]$\lambda 5007$ is the thick line and the dotted line
represents H$\beta$.\label{perfiles}}
\end{figure}

Using the IRAF task \textit{fit1d}, a ``continuum frame'' was made by
fitting 3 cubic spline pieces to every line in the spectra, and then,
a ``line frame'' was constructed by subtracting the ``continuum
frame'' from the original 2-D spectroscopic frame.  The spatial
distribution of the continuum is obtained by column-averaging the
whole ``continuuum frame,'' and the spatial distribution of the
emission lines are obtained by column averaging the ``line frame''
over the columns that span $1.25 \times$ the 
$\mathrm{FWHM}$ of the line.  The spatial profiles for the two new
objects studied are given in figure \ref{perfiles}.

Figure \ref{perfiles} shows that:

\begin{enumerate}

\item{The spatial distribution of the ionized gas and underlying
stellar population in object 1 is very similar to the distributions
observed in objects 3 and 4 \citep{hoyos04}: The nebular
component light centroid does not coincide with the continuum
centroid although the peaks of the different components clearly line up.
The two peaks observed at +1.5 kpc and -2.7 kpc in the H$\beta$ spatial 
profile are not real. 
This object is very similar to the cometary BCDGs.
In this particular case, the spread of the continuum is
about twice the size of the line-emitting region. In the optical
images, this object shows a conspicuous tail, and a very bright
knot. It is also seen that the spatial extent of the two emission
lines used to trace the ionized gas is very similar.}

\item{For object 2, the diferences in the centroid positions for the
continuum and the line-emitting region are very small when compared to
the spread of both. In this case, the ionized gas can be said to lie
in the center of the optical galaxy. This is to say that the line emitting region
coincides with the continuum light centroid. The true barycenter of these
galaxies is not known. It is also interesting to note
that, in this particular case, the distributions of each component are
very noisy even for the continuum, and that the distributions of
H$\beta$ and [\ion{O}{3}]$\lambda 5007$ are very similar to each
other.}

\end{enumerate}

As in \citet{hoyos04}, the H$\beta$ and [\ion{O}{3}]$\lambda 5007$
lines were chosen to probe the ionized gas phase because
[\ion{O}{3}]$\lambda 5007$ and H$\beta$ are very close to each other
in wavelength space and therefore they are affected by a similar
extinction. For this reason, differences in their spreads can then
reveal real differences in their distributions and not just a
reddening effect. As it was pointed out in \citet{hoyos04}, the
observed differences in the spread of H$\alpha$ and H$\beta$ or
[\ion{O}{3}]$\lambda 5007$ and [\ion{O}{2}]$\lambda 3727$ can be
accounted for by a moderate extinction, within the measurement errors,
making it impossible to detect extinction or excitation gradients
unambiguously. Finally, the inclusion of these two new objects to the
sample of LCBGs with high quality spatially resolved spectroscopy
allows us to give a first interval of the frequency of 
off-center\footnote{Here, the word ``center'' means the continuum
light centroid.} starbursts in LCBGs. 
Since it results that three out of six LCBGs with
STIS spectroscopy happen to have their starburst displaced from their
centers, all that can be said is that the fraction of LCBGs with
shifted starbursts is $50^{+49}_{-20}\%$. More observational work
is needed to asses the frequency and impact of this issue in the evolution
of LCBGs.

\subsection{Line Ratios, Reddening, Equivalent Widths and Metallicities.\label{numeros}}

The observed flux of each emission line and the equivalent widths have
been measured via gaussian fits using the STSDAS task
\textit{ngaussfit}, according to the method outlined in
\citet{hoyos04}.  This task can fit to a set of x--y values the sum of
a straight baseline, and one or several gaussian
profiles. This task also gives approximate errors for the fitted
parameters, calculated by resampling. The emission lines were fitted
rather than summed because of residuals of bad pixels.

Two different positions are considered for each galaxy corresponding
to (a) the inner and (b) the outer regions of each object. Despite
these names, both the (a) and (b) regions defined here are
\emph{within} the line-emitting region. These two one-dimensional
extractions are defined as to encompass approximately the same line
luminosity. It is in these two regions that the following
stellar population analysis is carried out.
These apertures are near one effective radius for the line-emitting region, as
detected through the spectroscopic aperture. The linear sizes of both
extractions are given in the second row of Table \ref{stix}, in kpc.  For
position (a), both the FWHM and central wavelength are left as free
parameters. However, for position (b), the FWHM is fixed to the FWHM
derived for position (a) to decrease the number of free parameters and
hence increase the goodness of the fit in poor S/N conditions.  The
[\ion{O}{2}]$\lambda 3727$, H$\beta$ and both [\ion{O}{3}] lines can
be observed in all cases. However, athough H$\alpha$ is detected for
object 4, it is affected by many bad-pixels that render the line
unusable.  In the case of object 1, H$\gamma$ is detected, too.

The amount of extinction is here parametrized using the logarithmic
extinction coefficient c(H$\beta$). It was derived using standard
nebular analysis techniques \citep{ost89} using the H$\alpha$/H$\beta$
Balmer line ratios.  For object 1, the H$\gamma$ line was used to
derive the extinction coefficient, which agreed well with that derived
from H$\alpha$ and H$\beta$ but had a smaller uncertainty. Since the
observed galaxies are at high Galactic latitude, suffering galactic
extinctions $A_{B}<0.11$ mag which corresponds to a c(H$\beta$)$<0.05$
\citep{bh82,santam74}, it was assumed that all the extinction occurs
within the target objects. 
This approximation is further validated by the fact that the incoming
light is redshifted and therefore the extinction will be much lower.
Consequently the spectra have been dereddened in the rest frame of the
targets according to a standard reddening curve \citep{whit58},
assuming case B recombination theorethical line ratios. The use of
different extinction curves would not significantly change the results
since they do not differ from the one used here in the wavelengths of
interest. The value of c(H$\beta$) is determined independently for the
inner and outer regions for objects all objects but H1-13088, for
which H$\alpha$ is unfortunately not well measured. A global value of
c(H$\beta)=0.1\pm0.05$, typical for local \ion{H}{2} galaxies
\citep{hoyos_diazxx} is adopted. No underlying stellar
hydrogen absortion is detected, as the continuum is very low.

Line equivalent widths were measured in the STIS spectra. It is possible
to measure the line equivalent widths despite the fact that
these spectra are emission-line dominated, if the error estimates are 
carefully estimated. This is explained in detail in \citet{hoyos04}, where it 
is also possible to see four examples of the extracted spectra 
\citep[see again][Fig 2.]{hoyos04}.

The measured extinction coefficients and de-reddened line ratios are
given in Table \ref{stix}, together with the measured \textit{rest-frame}
equivalent widths. It is seen that the extinction values are always
very small. It can also be seen that the ionization ratio 
$\log\mbox{[\ion{O}{3}]/[\ion{O}{2}]}$ and both the 
$\log\mbox{[\ion{O}{3}]/[\ion{H}{2}]}$ and 
$\log R_{23}$ numbers are always smaller in the outer zones of the observed galaxies than in the inner
part of the line-emitting region.
$R_{23}$ is defined as the reddening-corrected 
([\ion{O}{2}]$\lambda 3727$+[\ion{O}{3}]$\lambda ,\lambda 4959, 5007$)/H$\beta$ ratio.
In some cases, the differences are
larger than the derived errors revealing the existence of a
gradient. For some other cases the differences are more
dubious. However, even though it is not possible to claim that these
gradients have been detected for \emph{all} the individual objects
studied, the fact that these differences between the (a) and (b) zones
hold for the \emph{majority} of cases allows to think that these
gradients are real at least for typical LCBGs.

\subsection{The Oxygen Content. Metallicity Estimates.}

One of the main goals of any nebular analysis is to estimate the
metallicity of the emitting nebula. There are many reasons to
determine the metallicity of the ionized gas in a galaxy, but chief
among them is that it yields an up-to-date determination of the
chemical composition of the most recent generation of stars born in
the galaxy studied.

The emission line spectrum from a star forming region depends strongly 
on the electron temperature and metallicity of the nebula. 
The ratio of the auroral line [\ion{O}{3}]$\lambda 4363$ to the lower 
excitation lines [\ion{O}{3}]$\lambda \lambda 4959,5007$ gives a direct
determination of the electron temperature where O$^{+}$ and O$^{++}$ are  
the dominant species, but beyond certain metallicities, the increasing 
cooling leaves no energy to collisionally excite the upper levels.
When this is the case, one must resort to determine the oxygen
abundance empirically.

In principle, the oxygen content can be determined from the intensities
of the optical emission lines using the $R_{23}$ number, defined as
the reddening-corrected ratio ([\ion{O}{2}]$\lambda
3727$+[\ion{O}{3}]$\lambda ,\lambda 4959, 5007$)/H$\beta$. This method
was originally developed by \citealp{pag79} 
\citep[see also][among many others]{pes80,ep84,mcg91,dop02}.  The
main problem with this empirical method is that the calibration is
double-valued, and some \textit{a priori} knowledge of the metallicity
range is needed in order to solve this degeneracy. Since the S/N of
the spectra used is not very large, no other lines that could help
break this degeneracy are detected. Thus, the lack of this initial
knowledge about the oxygen content remains, and the metal content of
the galaxies studied cannot be cleanly determined.  Instead, the
metallicity values that result from the use of the \citet{py00}
calibration for both the lower and upper branches are calculated.  The
 metallicity estimates calculated using the \citet{py00} formulae
 are given in Table \ref{stix}.
Unfortunately, the measured values for $\log R_{23}$ squarely place
our objects in the turnaround region of the
$12+\log\mbox{(O/H)}$--$\log R_{23}$ plot. In this regime, the only
thing that can be safely said is that the metallicity is between 7.7
and 8.4. Therefore, the analysis can only safely conclude that these LCBGs have an
oxygen content greater than 7.7, but lower than the solar value of 8.69 \citep{alleprie01}
It is also worth mentioning that, with the available data, it is not
possible to unambiguosly detect any gradient in the metallicity of these
objects since the derived values of $R_{23}$ imply that the oxygen
content of both zones can lie in a very wide range\footnote{In
addition, the metallicity gradients of compact systems are likely to
be intrinsically small because of the mixing.}.  It can also be
concluded that the $R_{23}$ method is not best suited to determine the
gas-phase metallicity of LCBGs because the $R_{23}$ diagnostic is likely to
fall in this regime.
 
However, one initial metallicity estimate can be
provided by the ratio [\ion{N}{2}]$\lambda 6584$/H$\alpha$ as
suggested by \citet{dtt02}.
Since [\ion{N}{2}]$\lambda6584$ can not
be measured in the spectra analyzed here, as it is below
the detection limit due to poor S/N,
it is not possible to use the spectra used here directly for this
task. The only available information are the \citet{guz96} W.M. Keck Telescope
[\ion{N}{2}]$\lambda 6584$/H$\alpha$ measurements for a set of very
similar sources from the same parent sample.
These measurements consist of very high quality spectra obtained using the HIgh REsolution
Spectrograph (HIRES) The average oxygen content of the \citet{guz96}
sources, derived using the \citet{dtt02} 
expression for 12+log(O/H), is $12+\log \mathrm{(O/H)}=8.6\pm0.1$. This value is a little bit too high to be
completely compatible with the majority of the \citet{py00}
metallicity calculations presented above, although it is compatible
with some of them. This indicates that the oxygen content is much more
likely to lie in the upper half of the interval allowed by the oxygen
line ratios. It is interesting to note that the HIRES measurements
of [\ion{N}{2}]$\lambda 6584$/H$\alpha$ are all around 0.18, and that
the $3\sigma$ detection threshold at H$\alpha$ is around 15\% of H$\alpha$
for the STIS spectra, meaning that the STIS spectra were very close to detect
the nitrogen line. This is true for both the low resolution and high resolution
configurations.


\begin{deluxetable}{lllllllll}
\tabletypesize{\scriptsize}
\rotate
\tablewidth{0pt}
\tablecaption{Absorption Coefficients, De-reddened Line Ratios, Rest-Frame Equivalent Widths, Oxygen Content and Stellar Colors.
\label{stix}}
\tablehead{
\colhead{ID} & \colhead{1} & \colhead{1} & \colhead{2} & \colhead{2} & \colhead{3} & \colhead{3} & \colhead{4} & \colhead{4} 
} 
\startdata

Position.   &	 inner        &	 outer &	 inner        &	 outer        &	 inner       &	  outer        &	 inner         &	 outer          	\\
Size(kpc).     &	 0.83         &	 1.66  &	 1.04         &	 2.01         &	 1.28        &	  2.03         &	 1.27          &	 3.04           	\\
c(H$\beta$) &	 0.0$\pm$0.1  &	 0.0$\pm$0.1 &	 0.0$\pm$0.1  &	 0.0$\pm$0.2  &	 0$\pm$0.1   &	 0.10$\pm$0.2  &	 0.10$\pm$0.05 &	 0.10$\pm$0.05  	\\
$\log (($[\ion{O}{3}]$\lambda \lambda 4959,5007$)/H$\beta$     &	 0.71$\pm$0.04  &	 0.60$\pm$0.03     &	 0.60$\pm$0.03  &	  0.39$\pm$0.04  &	 0.67$\pm$0.03  &	 0.44$\pm$0.06  &	 0.7$\pm$0.1    &	 0.54$\pm$0.08  	\\
$\log($[\ion{O}{3}]$/$[\ion{O}{2}]$)$\tablenotemark{a}          &	 0.48$\pm$0.04  &	 0.38$\pm$0.02    &	 -0.31$\pm$0.04 &	  -0.38$\pm$0.05 &	 0.43$\pm$0.06  &	 0.33$\pm$0.09  &	 0.28$\pm$0.09 &	 0.26$\pm$0.07  	\\
$\log R_{23}$\tablenotemark{b}  &	 0.83$\pm$0.04  &	 0.75$\pm$0.02      &	 1.08$\pm$0.02  &	  0.92$\pm$0.06  &	 0.80$\pm$0.03  &	 0.61$\pm$0.05  &	 0.89$\pm$0.08 &	 0.73$\pm$0.06  	\\
$\log W_{[OII]\lambda 3727}$\tablenotemark{c}                &	 1.88 &	 1.94   &	 2.18    &	 2.11    &	 1.65  &	 1.43 &	 1.93    &	 1.79    	\\
$\log W_{H\beta}$\tablenotemark{c}                                 &	 1.79 &	 1.87   &	 0.60    &	 0.47    &	 1.59  &	 1.51 &	 1.67    &	 1.57    	\\
$\log W_{[OIII]\lambda \lambda 4959+5007}$\tablenotemark{c} &	 2.51 &	 2.53   &	 1.10    &	 0.77    &	 2.28  &	 2.00 &	 2.40    &	 2.13    	\\
$\log W_{H\alpha}$\tablenotemark{c}                                  &	 2.00 &	 1.82   &	 \nodata &	 \nodata &	 2.30  &	 2.16 &	 \nodata &	 \nodata 	\\
NICMOS $F_{\lambda}$\tablenotemark{d}  &   $2.8\pm0.3$ &   \nodata      &  $5.9\pm0.6$   &  \nodata  &  $3.9\pm0.4$  &   \nodata & $10\pm1$  &  \nodata \\      
NICMOS Luminosity \tablenotemark{d}  &   2.7   &   \nodata      &  3.9   &  \nodata  &  2.4  &   \nodata & 7.0  &  \nodata \\      
$12+\log\mathrm{(O/H)}_{\mbox{hi}}$\tablenotemark{e} &	 8.4 &	 8.5   &	 7.8 &	 8.1 &	 8.4 &	 8.7 &	 8.2 &	 8.5 	\\
$12+\log\mathrm{(O/H)}_{\mbox{lo}}$\tablenotemark{f} &	 7.8 &	 7.7 &	 8.8 &	 8.6 &	 7.7 &	 7.5 &	 8.0 &	 7.7 	\\
$\bv_{\mbox{Observed.}}$\tablenotemark{g} &	   0.41 &	 0.49   &	 0.28   &	 0.27  &	 0.08     &	 0.25    &	 0.26   &	 0.35   	\\
$\bv_{\mbox{Stellar.}} $\tablenotemark{h} &	   0.13 &	 0.20   &	 0.34   &	 0.33  &	 -0.08    &	 0.16    &	 0.12   &	 0.28   	\\

\enddata
\tablenotetext{a}{This is defined as the reddening-corrected ratio 
[\ion{O}{3}]$\lambda ,\lambda 4959, 5007$/[\ion{O}{2}]$\lambda 3727$. The other line ratios are also reddening-corrected.}
\tablenotetext{b}{$R_{23}$ is defined as the reddening-corrected ratio 
([\ion{O}{2}]$\lambda 3727$+[\ion{O}{3}]$\lambda ,\lambda 4959, 5007$)/H$\beta$.}
\tablenotetext{c}{The rest-frame equivalent widths are in \AA \ . The average error in the 
rest frame equivalent width is 17\%. This corresponds to an error of 0.07 in the log. Most errors 
fell in the 15\%-20\% interval. The G750M spectra used for object 2 bear the largest uncertainties 
since the slit width used was 0.2\arcsec wide, compared to the 0.5\arcsec slit used for the other 
three objects. The errors in the equivalent widths for object 2 are therefore 30\%, corresponding 
to 0.13 in the log.}
\tablenotetext{d}{Units of $F_{\lambda}$ are $10^{-15} $erg s$^{-1}$ cm$^{-2}$. The reported luminosities are rest-frame luminosities in the \emph{rest-frame} wavelength range that shifts to the F160W wavelength passband. The luminosity units are $10^{42}$ erg s$^{-1}$, and the errors are typically 15\%. These numbers both refer to the \emph{total} luminosity from the line-emitting region.}
\tablenotetext{e}{Oxygen abundance derived using the calibration for the upper branch \citep{py00}. The solar value is 8.69 \citep{alleprie01}.}
\tablenotetext{f}{Oxygen abundance derived using the calibration for the lower branch \citep{py00}. The solar value is 8.69 \citep{alleprie01}.}
\tablenotetext{g}{\emph{Observed} colors. The uncertainty is 0.01 magnitudes in all cases.}
\tablenotetext{h}{\textit{Rest-frame}, \emph{stellar} population only, extinction corrected $\bv$ colors.
This is defined in the text. The uncertainty is 0.09 magnitudes.}

\tablecomments{In the second row the physical sizes of each aperture in linear kpc are given.
For object 1, the logarithmic extinction coefficient was derived from the $H\gamma/H\beta$ ratio. Its value
was consistent with the value derived using $H\alpha$, but the error was smaller.
For object 2, the [\ion{O}{2}]$\lambda 3727$ flux is first measured in the integrated 
G430L spectra. The intensities in each zone (a or b) are then derived by requiring 
that the zones emit 50\% of the light each.
This is true to within 5\% for the other three lines seen in the 
other instrumental setup. The reported values of $\log W_{([OII]\lambda 3727)}$ correspond to 
the integrated spectra too.
For object 4, H$\alpha$ is not observed and the listed value for the logarithmic extinction 
coefficient is an \textit{assumed} value, typical for local \ion{H}{2} galaxies.}
\end{deluxetable}


\subsection{Scaling and Registering of the Data. Calculation of Stellar Colors and NIR Luminosities.}

In order to describe the stellar populations found in the galaxies
studied, it is needed to match the STIS spectroscopic frames with the
WFPC-2 images and the NICMOS frames. 
This is done by forcing that each galaxy's centroid in
the WFPC2/NICMOS images coincides with the center of the continuum
distribution of the STIS spectroscopic frames. Once both centroids
are calculated, the slit orientation with respect to the image is used
to determine what pixels in the WF/PC-2 frames correspond to the
\textit{inner} and \textit{outer} zones defined previously for the
STIS spectra. 

\begin{figure}
\plotone{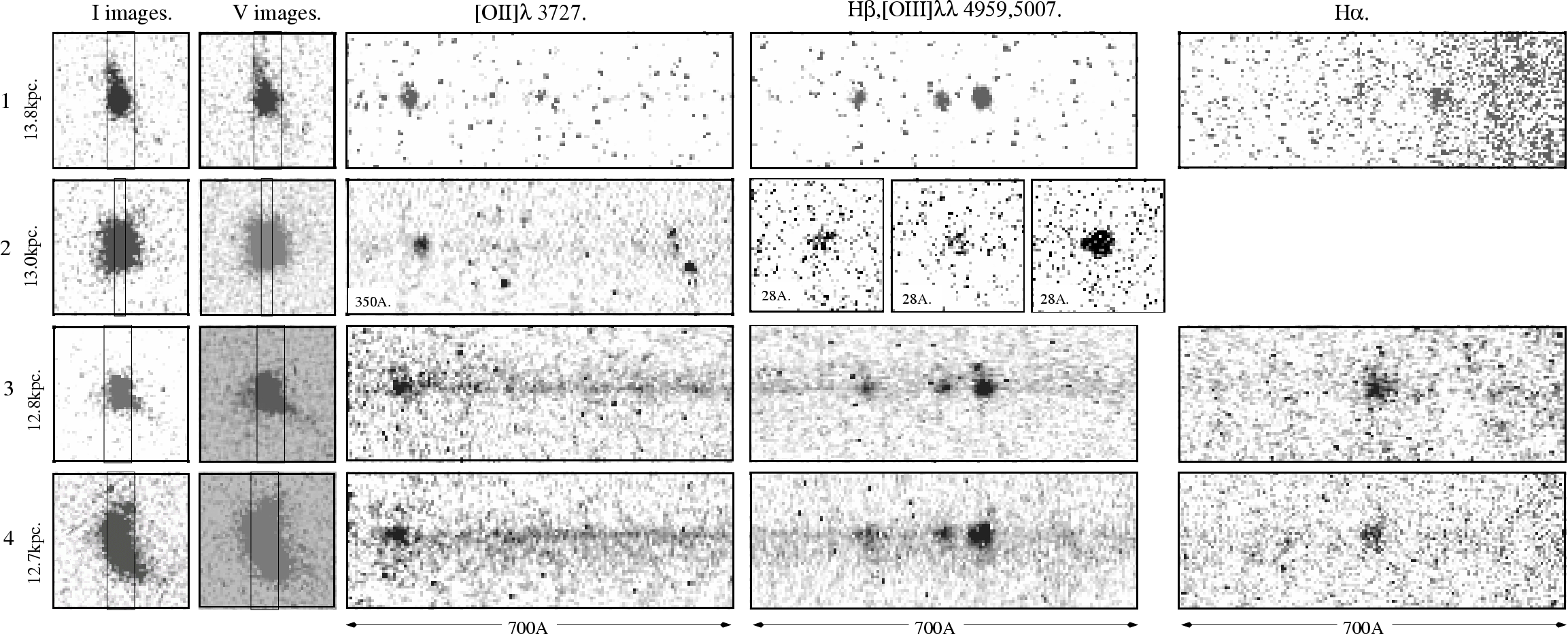}
\caption{Matching between the STIS 2-D spectra and the HST/WFPC-2
images. Each spectral strip spans 2.5\arcsec in the spatial
direction. The linear scale is given for each object in the
figure. Each spectral strip spans 700\AA {} except for object \#2, in
which the wavelenght coverage is given in each image. The spectral strips 
presented for this objects are not registered with respect to the spectral
strips shown for the other objects for this reason. The H$\alpha$ strip
for object 1 is not perfectly registered with respect to the H$\alpha$ strips
for objects 3 \& 4 because the H$\alpha$ line fell very near to the CCD edge.
This spectral strip still spans 700\AA, though.
The slit orientation is shown superimposed on HST F606W and F814W images which
are 2.5\arcsec$\times$2.5\arcsec on a side.
It is interesting to note that the spatial distribution of the continuum
is lopsided with respect to the peak. This is most evident in objects 3 \& 4. \label{registro}}
\end{figure}

Figure \ref{registro} shows the slit placement in the F814W and F606W
WF/PC-2 images. The bidimensional spectra presented for objects 1, 3
\& 4 are low-resolution G750L frames, whilst the spectrum presented
for object 2 is a high-resolution G750M frame. In this figure, the
spectra have been scaled and approximately aligned with the images.
It is seen that the STIS spectra only probe the most luminous regions
of the observed objects, corresponding to the line-emitting
region. This area spans less than $1\arcsec$, and it's been further divided
into two regions. This makes it neccessary the use of the spatial resolution of the HST
instruments. Even though these objects have been observed from the ground in the $UBRIK$ bands,
it is not possible to merge those data with the observations presented here because the spatial
resolution of those observations is not high enough to study the presented objects at the subarcsecond level.
 The stellar population analysis presented in the following
section \ref{calcus} is therefore best suited for the volume within
the line-emittion regions of the studied galaxies. This region
typically has a radius of 1.7 kpc.

\begin{figure}
\plotone{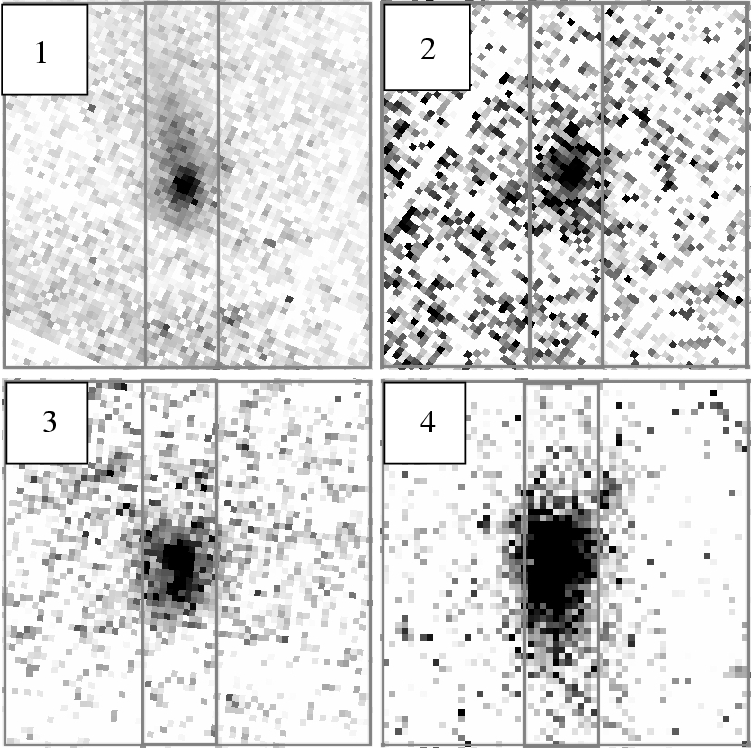}
\caption{NICMOS F160W reduced images. Each postage stamp
is $2.5\arcsec$ on a side. The slit is placed as in Figure \ref{registro}.
The number in the upper left corner of each panel corresponds to the identifying numeral
of Table \ref{prestab}.
 \label{putonicmos}}
\end{figure}

Figure \ref{putonicmos} shows the NICMOS frames. Each panel is $2.5\arcsec$ on a side.
Again, the images have been rotated so that the slit silhouette
is vertical, as shown in the figure. It is seen that
the STIS spectra focus on the very central regions of the
objects.

The next step is to determine the $\bv$ color of the areas in the
WF-PC-2 F814W and F606W images that match the inner (a) and outer (b)
zones of the STIS spectra. In order to accurately determine said
color, several issues are to be taken into account.  The first of them
is to correct the integrated DN (CCD counts) in each area for the imperfections of
the charge transfer mechanism. There are several choices to accomplish this task.
The version used in the WFPC2 Instrument Handbook, based on the 05/31/2002
version of Andrew Dolphin's calculations was used.
This correction depends on the objects' location in the CCD. It also depends
on the objects' accumulated DN and CCD gain. This correction is obviously insensitive
to the filter used. Fortunately, for the observations used, any given galaxy
is always imaged in the same CCD location, and the accumulated counts are very similar 
for the two filters used. This correction is therefore very small. It only changes
the observed colors by less than 0.1 magnitudes, usually much less. The statistical
error introduced by this correction was estimated by varying the DN parameter
of the software tool previously mentioned around the observed DN number. It turned
out to be that the magnitude error introduced by this correction
was negligibly small, 0.001 magnitudes in the worst case.
For a complete review of this problem and the different approaches to 
correct for it, see \citet{dolf00}.

The second issue is the derivation of the \textit{k}-correction.
This step is needed to convert from the \emph{observed} $I-V$ colors
to \emph{rest-frame} \bv colors. The fits 
given in \citet{fuku95} for the case of an Irregular Magallanic spectrum are used.
The \textit{k}-corrections are 0.32 for objects 3 \& 4, 0.39 for object 2 and
0.47 for the furthest object. The errors introduced by this correction
are mainly systematic, arising from deviations of the real galaxy spectra from
the assumed Irr spectrum. These are estimated in around 0.08 magnitudes, as
was done in \citet{guz98}.

The third issue is to derive the \textit{emission-line free} colors.
The method presented in \citet{vri_telles97} is adopted. 
This method uses the \emph{rest-frame} equivalent widths, together with
the $B$ and $V$ filter transfer functions to estimate the contribution
of the emission lines to the observed galaxy colors.
The method is now sketched.
Given any \emph{observed} color $A-B$, it can be written as:

\begin{equation}
B-A=C_{BA}-2.5\times \log \frac{L^{c}_{B}+L^{l}_{B}}{L^{c}_{A}+L^{l}_{A}}
\end{equation}

\noindent
where $C_{BA}$ is the photometric constant, $L^{c}_{X}$ is the continuum luminosity
in the $X$ band and $L^{l}_{X}$ is the line luminosity falling within the $X$ band.
\indent

If the line luminosities are considered as small perturbations to the spectrum, the above equation can be cast
into the following form.

\begin{equation}
B-A=(B-A)_{0}+1.086\times (L^{l}_{A}/L^{c}_{A}-L^{l}_{B}/L^{c}_{B}) 
\end{equation}
\noindent
where $(B-A)_{0}$ is the color that would be observed in the absence of the emission lines.
\indent

It is also possible to write the color correction as:

\begin{equation}
A-B=(A-B)_{0}+1.086\times \left(\frac{L^{l}_{A}}{I_{A} \times W_{A}}-\frac{L^{l}_{B}}{I_{B} \times W_{B}}\right) 
\end{equation}
\noindent
where $W_{X}$ is the bandpass of the $X$ band, and $I_{X}$ is the average specific intensity.
\indent

The color correction can therefore be written as:

\begin{equation}
1.086\times \left(\frac{ \sum A_{i}\mbox{EW}(i)}{W_{A}}-\frac{\sum B_{i}\mbox{EW}(i) }{W_{B}}  \right)  
\end{equation}
\noindent
where $\mbox{EW}(i)$ is the line equivalent width of the $i^{th}$ emission line, and $X_{i}$ is the filter transfer function
evaluated at that wavelength.
\indent

It is important to note that, even if the line equivalent widths used in the above formulae bear
large uncertainties, any error in the measured equivalent widths is greatly diminished or
cancelled out. This happens because the same equivalent width appears in both terms of the 
color correction with opposite signs, although with possibly different multiplicative factors.

The applied corrections range from essentially zero to 0.25 magnitudes. The statistical
error introduced by this correction is very small. Given the derived
uncertainties in the measured equivalent widths, these errors are less than
0.02 magnitudes.

Finally, the E($\bv$) color excess is derived from the c(H$\beta$) values 
for each zone. These color excesses are very small and are virtually 
error free.  The resulting colors are
therefore \textit{rest-frame}, \emph{stellar} population only,
extinction corrected \bv colors.
The total photometric errors, obtained adding in quadrature all the
aforementioned uncertainties, are estimated in 0.09, for all objects. These
colors are gathered in Table \ref{stix} for the inner and outer parts
of the line-emitting region in which the stellar populations are
studied.

The NICMOS data were used to calculate the \emph{total} F160W luminosities
of the line emitting regions. It is needed to use the NICMOS data in this way 
because of the larger angular extent of the NICMOS Point Spread Function, that makes it
impossible to calculate the observed fluxes of the (a) and (b) zones separately
with the same accuracy as for the optical frames. 
The measured fluxes and luminosities are gathered in Table \ref{stix}.
The reported luminosities all assume $c(H\beta)=0.2$, and they also
assume that $A_{X}/A_{V}=0.4$, where $A_{X}$ is the extinction in the
wavelength range that maps into the F160W passband. This value is the average value
for that ratio between the $I$ and $J$ bands. The contribution from the
nebular continuum, which begins to be important at these wavelengths
has been estimated at less than or around 10\% using the 
standard techniques presented in \citet{ost89}.

It is seen in Table \ref{stix} that the colors of the inner zones are
usually bluer than those of the outer zones of the line emitting
region, in accordance with the \citet{guz98} color profiles.
This can be interpreted in the light of the results presented
in subsection \ref{numeros}.
Whenever one region in a galaxy is redder than other region, it can be
caused by differences in reddening, metallicity, or age. The first of
these factors is unlikely to play a role in this particular case since
the reddening has been estimated, and it is very small. The second
factor is also unlikely to be responsible of the observed color
gradients since the metallicity differences are likely to be very
small in these systems at such short spatial scales because of very efficient
mixing by supernova driven winds. It is moreover concluded
that the observed color differences between the (a) and (b) zones are
more likely to arise from age differences between these two
zones. Specifically, the underlying stellar populations in the (b) zones 
are typically older than the stellar populations of the (a) zones of their respective
galaxies. The chemical enrichment histories of both regions have probably been
very similar for the reasons outlined above.
This is an important finding for these distant and compact sources.

Alternatively, it might be possible to 
think that the color gradient is caused by mass segregation. If , for instance, most of the newly formed ionizing 
stars were in the (a) zone, the color of the inner part would be naturally bluer than the color of the
outer zone. This scenario could also explain the observed differences in the ionization ratio 
between the two zones, which would be higher in the inner zone because
of the higher amount of ionizing photons available to ionize the inner parts of the nebula.
Although mass segregation is a concern for massive clusters
(see, for example \citet{stolte05} and \citet{grijs02}), it only operates at very short distances (10 pc at most), and
not at the much larger scales of interest in the present case. This possibility 
of a spatially varying Present-Day Mass Function can then be safely ruled out.


\section{The Evolutionary Models.}
\label{calcus}

The observational data gathered for the LCBGs presented here can be used to construct galaxy models
that represent the objects. Although the models presented are very crude and tailored, they provide at least
an idea of the stellar inventory and possible evolution of these LCBGs.
This will shed light on the possible evolution and age of these galaxies. In this section, the star 
formation history of the four LCBGs is studied using evolutionary population synthesis techniques.
The models have been built using the \citet{bruzual03} and \citet{sb99} 
evolutionary models. For each galaxy \emph{and extraction, (a) or (b)}, two simple stellar populations are combined.
One population is designed to mimick the observed properties of the ionizing stellar population, and
the other population is chosen to match the observed colors when combined with the ionizing population.
This latter population therefore represents the underlying population.
We note that the population analysis which will be used here is carried out for
both the inner and outer zones of the line-emitting region \emph{separately}, taking
advantage of the HST spatial resolution. In addition, the fact that the star forming region
is split into two subregions can also help to model possible delays
between the star formation episodes in the inner zone and in the outer zone of the
line-emitting region.

The metal content of the ionizing stellar population is assumed to be equal to the metal content
of the ionized gas. Since the only information available on the metal content of the ionizing population
of these sources comes from the gas phase oxygen 
abundances\footnote{12+$\log$(O/H) lies between 7.8 and 8.4, although the preferred value is 8.4
because of the \citet{guz96} HIRES measurements.}, two sets of models are constructed. 
Models \#1 assume that the ionizing population has a metallicity of $Z=0.008$. Models \#2 adopt $Z=0.004$ for
the metal content of the ionizing population. These two numbers can be mapped into
$12+\log\mathrm{(O/H)}=8.4$ and $12+\log\mathrm{(O/H)}=8.0$, approximately, assuming solar 
chemical proportions. Therefore, for each galaxy and extraction, two models of different metallicities are built.
The stellar inventory of the observed galaxies is described according the following steps:

\begin{enumerate}
\item{The first step consists on the derivation of the masses of the ionizing populations.}
\item{The second step is the determination of the ages and colors of the ionizing populations. In this step
the \cite{sb99} models are used.}
\item{In the third step, the mass of the underlying stellar population is derived, together with 
its age. The \citet{bruzual03} models are used to represent the older stellar generation.}
\item{The last step is a double check of the derived masses using the measured NICMOS luminosities.}
\end{enumerate}

\subsection{The mass of the ionizing populations.}

The mass of the ionizing population is a very important parameter that defines it. 
If the mass of the ionizing generation is high enough, the initial mass function will be well sampled.
If the initial mass function is not well sampled, it might be the case that the models can not be used
to describe the stellar population. This effect is commonly known as the \emph{richness effect}
or also as the \emph{poisson shot noise}.
Depending on the slope and cutoff masses of the IMF used, this can become a problem for population masses lower 
than $10^{5} M_{\odot}$, but for the usual Salpeter IMF the mass limit is lower.
This issue has been extensively studied in the series of papers \citet{cervino00,cervino01,cervino02}.

The mass of the ionizing population, M$_{*}$ can, in principle, be derived from the luminosities of
the observed recombination lines according to the following
considerations.
As the stellar population ages, the number of hydrogen ionizing photons per unit
mass of the ionizing population also decreases. Assuming that the age of the
ionizing population is related to the intrinsic W$_{H\beta}$, a relation should exist
between the number of hydrogen ionizing photons per unit
mass of the single stellar generation and W$_{\beta}$. Such relation is given for
single-burst models in \citet{diazmodelsXX}. The expression used is

\begin{equation}
\log(Q(H)/M_{*})=44.8+0.86\times\log W_{H\beta}
\label{angeles}
\end{equation}
\noindent
where Q(H) is the number of hydrogen ionizing photons per second, M$_{*}$
is expressed in solar masses, and $W_{H\beta}$ is the ``intrinsic'' equivalent width, the
one that would be observed if no underlying stellar population was present. Again, this expression
works best in stellar populations with well sampled IMFs. The number of ionizing photons
is an integral over the mass range more affected by the richness effect, and it is therefore
very sensitive to small number statistics when applied to low mass populations.
The IMF used to derive this expression had an exponent of 2.35, a a lower mass limit of 0.85$M_{\odot}$
and a upper mass limit of 120 M$\odot$. 

\indent

However, the presence of the underlying stellar population introduces a degeneracy in the measurements
of the equivalent widths. This prevents us from measuring directly the equivalent widths that would be observed
if only the ionizing stellar population was present. The main goal of the stellar population analysis presented
here is then to unravel the different stellar populations. This is carried out following the ideas presented 
in \citet{diaz00}. In order to solve this degeneracy in
equivalent width introduced by the presence of the underlying population, the 
upper envelope of the relationship between $\log W_{\mathrm{H}\beta}$ and $\log($[OIII]$/$[OII]$)$ 
presented  in \cite[Figure 6 of]{hoyos_diazxx} was used to estimate the equivalent
width that would be observed in the absence of any previous stellar 
generation. This ``effective'' equivalent width is called $W^{0}_{\beta}$.
This assumption is now briefly justified.
In the W$_{H\beta}$--[\ion{O}{3}]$\lambda \lambda 4959+5007$/[\ion{O}{2}]$\lambda 3727$ ratio diagram
for very luminous \ion{H}{2} galaxies, if the observed range of ionization degrees is ascribed
to the age of the ionizing population, it is reasonable to assume that 
the vertical scatter shown by the data is caused by different contributions of continuum light from 
the underlying population\footnote{There is, however, a metallicity effect. In the
absence of any underlying stellar population, \ion{H}{2} galaxies with higher 
metalicity clouds will present a lower ionization ratio due to lower
effective temperatures of their ionizing stars.
However, there will be no change in the equivalent width of H$\beta$, to 
zero-th order. Therefore, $W^{0}_{\beta}$ should be a 
function of both the ionization ratio and metallicity. Only the 
ionization ratio dependence of the ``effective'' equivalent width will be used here.}.
The expression used is:

\begin{equation}
\log W^{0}_{eff}=2.00 \pm 0.02 + (0.703 \pm 0.028) \times\log \mathrm{[OIII]/[OII]}
\end{equation}

The upper envelope of this relation, which defines the ``effective'' equivalent width, is then 
relevant to estimate the ``intrinsic'' equivalent width 
because the objects located on top of this upper envelope are likely to 
be the ones in which the underlying population contribution is minimum.
This ``effective'' equivalent width allows us to estimate the mass of the ionizing population 
assuming that $W^{0}_{\beta}$ is equal to the real or ``intrinsic'' equivalent width
that would be measured if the starburst could be observed in
isolation. This is done via the expression given above using the measured H$\beta$ luminosities
to obtain the number of hydrogen ionizing photons per second through standard nebular analysis
techniques \cite{ost89}. The resulting stellar masses of the ionizing population
are given in Table \ref{res_1}.
Typical ionizing population masses range from 8 to 30 million solar masses. The systematic
errors affecting these calculations are explained in the following paragraphs.

The ``effective'' equivalent widths, which are mostly in the 100\AA -- 200\AA interval are
 still  lower limits only to the real value of the equivalent width
because the empirical method used here to derive the former does not forbid the existence of
objects with higher \emph{observed} equivalent widths for their \emph{observed} ionization ratio
[\ion{O}{3}]/[\ion{O}{2}]. This is indeed a possibility since the work presented in \cite{hoyos_diazxx} used
local galaxies and the galaxies analyzed here are intermediate-\textit{redshift} sources.
Therefore, the derived value for the mass of the ionizing populations obtained using
$Q(H)$ (or, rather, the H$\beta$ luminosity) and $W^{0}_{\beta}$ can only be only lower limits to the real masses.
It has to be said, however, that the possible differences between the ``intrinsic'' 
and the derived ``effective'' equivalent widths can not be very high. The maximum attainable
equivalent width, which corresponds to the case in which only the \emph{nebular} continuum is considered,
is 1000\AA. When the stellar continuum is taken into account, the maximum equivalent width then drops to about 400\AA, depending
on the IMF, metallicity and age of the population, and the largest $H\beta$ equivalent widths ever measured
are close to this latter quantity. This implies that the ``intrinsic'' 
equivalent widths of the observed starbursts have to be in the 200\AA -- 400\AA range.
These considerations allow us to estimate the possible systematic errors introduced by this approximation.
Using the typical ``effective'' equivalent width, and an average value for the maximum ``intrinsic'' equivalent
width, the real masses of the ionizing populations could be up to a factor of 2 larger.
However, even if the ``intrinsic'' value could be observed or somehow derived, the mass of the ionizing 
component obtained would always remain a lower limit because 
some photons could actually be escaping from the nebula, and dust globules will always be present amidst
the gas clouds. Moreover, if the ionizing population turned out to emit more photons harder than 1 Rydberg
than the number predicted by the models used to derive Eq. \ref{angeles}, the mass would be underestimated, too.
The new Geneva models with rotation at low metallicity \citep{maeder03} point towards this direction. For these 
reasons, further steps towards the determination of the ``intrinsic'' equivalent width from the ``effective'' one 
were not taken and both quantities are assumed to be equal.

\subsection{Determination of the Age and Intrinsic Color of the Ionizing Population.}

The value of the ``effective'' equivalent width can be used to obtain more information about the 
underlying stellar populations via the \cite{sb99} models, making some assumptions on the metallicity
and IMF of this stellar generation. Again, the ``effective'' equivalent width allows us to estimate, or 
at least to place significant lower limits to, the age and intrinsic color of the ionizing population
assuming that $W^{0}_{\beta}$ is equal to the real or ``intrinsic'' equivalent width
that would be measured if the starburst could be observed in isolation. The \citet{sb99} models
are used now.

\begin{figure}
\plotone{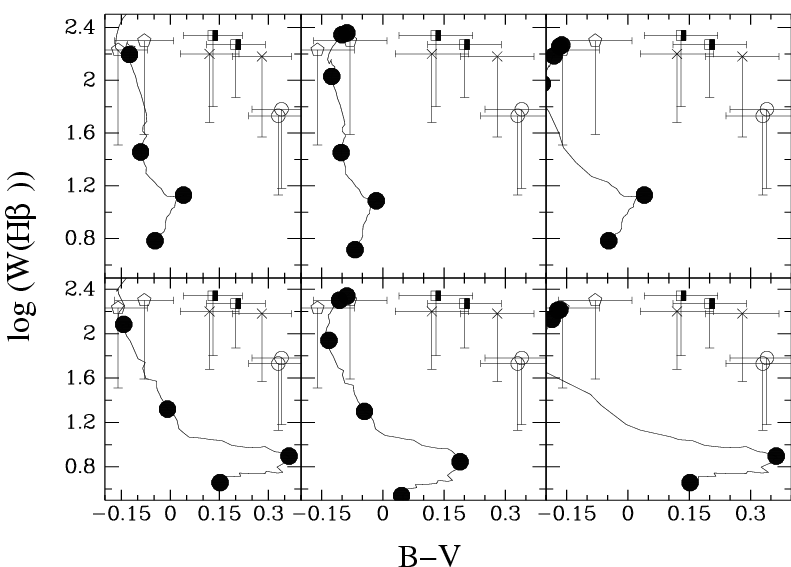}
\caption{Evolution of a single stellar population in the $\log W_{H\beta}$--$\bv$ plane from $t=0$Myr to $t=10$Myr. The data were taken
from the \citet{sb99} library. In all cases, an instantaneous burst is assumed, and the nebular continuum emission
is taken into account. Two different oxygen contents ($Z=0.008$, models \#1, lower panels, and $Z=0.004$, models \#2, upper panels) and three different IMFs (Salpeter from 1.0$M_{\odot}$ to 100$M_{\odot}$, left panels, a power law with an exponent $\alpha=3.3$ with the same mass limits, central panels, and a Salpeter IMF truncated at 30 $M_{\odot}$, right panels)  are plotted. For the LCBG points, $W^{0}_{\beta}$ is plotted against the \emph{emission-line free, reddening corrected} color. The horizontal error bars represent the 0.09 magnitude uncertainty, and the vertical semi error bar reaches till the \emph{observed} equivalent width, in order to indicate the importance of the contribution of the underlying stellar population to the optical light. The big black dots represent the 0,2,4,6,8,10$\times 10^{6}$ years checkpoints. Object 1 is represented by half-filled squares, object 2 is represented by $\bigcirc$, the location of object 3 is given by pentagons and $\times$ denotes object 4. Two points are given for each object, corresponding to the (a) and (b) extractions. In all cases, the (b) apertures have redder colors. \label{sb99_ionizing}}
\end{figure}

Figure \ref{sb99_ionizing} presents nine models from the Starburst99 \citep{sb99} library, which 
represent the evolution in the $\log W_{\beta}$ \textsl{vs.}\bv plane 
of a single stellar population from its birth till it is 10Myr old. The nebular continuum emission
is included in these diagrams. This evolution
is the solid line in those plots. Ages increase from top to bottom. Three different 
metallicities and IMFs are presented, as indicated in the figure. 
The derived $W^{0}_{\beta}$ for both the inner and outer zones of the line-emitting
 regions of the observed sources is plotted against the 
\emph{emission-line free, reddening-corrected} color in the case
of the studied LCBGs. This is the color that would be measured if 
the stars could be observed in isolation.

It is interesting to note that the values of the ``effective'' equivalent
width are close to the maximum equivalent width value of the plotted curves, which is the
maximum value of the ``intrinsic'' equivalent width for zero age clusters of the corresponding 
metallicity and IMF\footnote{About 350\AA, for the presented cases.} further highlighting the 
fact that, even though the ``effective'' equivalent width is a lower limit to the real value, it
is a reasonable approximation to it, specially if one considers that, even if the real value could be somehow
known, it would still yield lower limits to the mass of the ionizing clusters because
of photon escape and dust issues.
It is clearly seen that 3/4 of the regions studied require the presence of an 
underlying stellar population since they do not lie along any solid line. They can not
be reproduced by a single stellar population, then.
It is also seen that the single stellar populations with a truncated Salpeter IMF
have maximum ``intrinsic'' equivalent widths which are  \emph{lower}
than the derived values of $W^{0}_{\beta}$ of many regions. This means that we can rule out these models
as descriptions of the ionizing stellar populations found in the observed LCBGs since
the ``effective'' equivalent widths are in any case only a lower limit to the real one.
This can be interpreted in the sense that the observed galaxies have ionizing populations so massive
that they fully sample their respective IMFs, which is a necessary condition to achieve high equivalent widths.
In addition, the models whose IMF has a slope of 3.33 are redder than the inner region of object 3, which 
makes it impossible to accomodate any older and redder underlying population in this model
and there are also some regions with $W^{0}_{\beta}$ greater than the zero age equivalent width of
those models.
The models with Salpeter IMFs are therefore chosen to describe the ionizing populations, simply 
because they are compatible with all these restrictions.
Although the reasons to discard the $m^{-3.33}$ IMF are of course weaker, considering
multiple IMFs is by no means justified given the few inputs data at our disposal.
The use of the Salpeter IMF provides a crude yet \emph{consistent} description of the ionizing 
population found in the observed regions.

Assuming that $W^{0}_{\beta}$ equals the real equivalent width, as was done before, and
using the Salpeter IMF for the two metallicities that correspond to the metallicities
of the two models \#1 and \#2 used, it is possible to calculate the ages and intrinsic colors 
of the ionizing population. They are given in Table \ref{res_1}. The ionizing 
clusters are found to be very young (less than 5Myr in most cases).
As before, the use of the ``effective'' equivalent width instead of the unknown ``intrinsic'' value
introduces a systematic error in the age determinations. The ionizing populations are then
bound to be even younger than the presented quantities. When dealing with
very young coeval single stellar populations, it has to be kept in mind that
the observed properties of such populations do not change to a great extent until
they are about 3Myr old, simply because of the fact that this is the minimum lifetime
of even the most massive stars and also because the approximation of an \emph{instantaneous} burst
might fail at such short timescales. The main conclusion is that the ionizing populations of the observed regions
are probably extremely young.
The systematic error in the intrinsic color of the ionizing population is also estimated from the
color scatter of the models used between zero age and the maximum age allowed by the observed
$W^{0}_{\beta}$. All errors are 0.04 magnitudes.

\subsection{The Age and Mass of the Underlying Population.}

The next step to describe the stellar populations found in the
observed LCBGs is then to calculate the age and initial mass of
the underlying stellar population.
The single stellar populations used to describe the underlying
generation of stars are chosen from the \citet{bruzual03} library. The
evolutionary tracks used are the Padova 1994 ones and the IMF chosen
is the Salpeter IMF. The metallicity of this underlying stellar
population is $Z=0.004$ for the case of models 1 and $Z=0.0004$ for
the case of models 2. These are lower than the metallicities used
to describe the ionizing populations.
In this step, the ionizing cluster is represented by the single stellar population
available in the \citet{bruzual03} library that most closely resembles
the \citet{sb99} model used before in terms of metallicity, IMF and
$W^{0}_{\beta}$-derived age. 

Assuming that the metallicities of the ionizing populations
are higher than the metallicities of the underlying populations tries 
to emulate the chemical evolution of
the observed galaxies since their inception till the onset of the observed
starbursts. Although the choice adopted is not justified by the use of
any particular chemical evolution model, it is at least a zero-th
order solution to the problem since the \citet{bruzual03} models only
offer a limited array of metallicities for the different single
stellar populations.  In the framework of the closed box model and the
instantaneous recycling approximation, the corresponding
gas consumption fractions for the required metal enrichments are 10\% for 
models \#1 and 60\% for models \#2, assuming zero metallicity 
in the distant past, before the starbursts that gave birth to the underlying 
stellar generations took place.
The gas consumption fraction is larger in the metal poor models since 
the difference in metallicity between the two simple populations
mixed to construct them ($Z_{\mathrm{ion}}=0.004$, $Z_{\mathrm{underl}}=0.0004$)  is much 
larger than the metallicity difference of the two simple populations combined 
for the first models ($Z_{\mathrm{ion}}=0.008$, $Z_{\mathrm{underl}}=0.004$).
The yields provided by \citet{molla06_private} and \citet{maeder92} were used
in these latter calculations. 
Although these estimates, by themselves, do not prove that this picture
is accurate to describe the star formation history of these objects, they 
show that the assumed metallicity differences between
the stellar populations used to build thew models are, at least, possible.
In addition, it has to be kept in mind that the \emph{global} remaining 
gas fraction is bound to be lower for the metal rich models in the 
framework of the closed box model and the instantaneous 
recycling approximation than for the metal poor ones. 
In particular, given the assumed metallicity for the ionizing population
of the second set of models, the gas consumption fractions of \emph{pristine}
gass have had to be higher than 60\% in both cases, again in the closed box with
instantaneous recycling approximation.

The age and mass of the second stellar population are then chosen
so that the predicted equivalent width by the time the current
starburst is observed matches the \emph{measured} equivalent width and
the \emph{total} color equals the calculated \textit{rest-frame},
stellar population only, extinction corrected \bv color.
In essence, this procedure estimates the amount of ``red light''
that has to be added to the ``blue light'' of the ionizing population
in order to match the two observables available (\bv color and \emph{measured}
H$\beta$ equivalent width) at the time in which the observed starburst
takes place.This second stellar population is therefore designed to match
the older stellar population.

The difference between the \emph{observed} \bv color
and the previously derived \emph{intrinsic} \bv color of the 
ionizing population (given in Table \ref{res_1}) can be approximately
matched with ``how red'', and the difference between the \emph{observed}
equivalent width and the ``intrinsic''equivalent width roughly
translates into ``how much'' red light has to be added, although both
quantities are of course interlaced.

The errors in the observed equivalent widths (typically 17\%) and 
\emph{stellar} colors (0.09 magnitudes, as estimated above). Translate
themselves into errors in the relative mass of the underlying stellar population
with respect to the ionizing population and age. Although it is not true 100\%
of the error in the relative mass comes from the uncertainties in the
measured equivalent width and that 100\% of the errors in the age of the 
underlying stellar population arise from the errors in the derived
\emph{stellar} colors\footnote{In the sense that, for instance, iso-age contours 
in the age-relative mass plane do not map directly into iso-W$_{\beta}$ contours
in the W$_{\beta}$-\emph{stellar} color plane.}, these are the main
dependencies. Assuming this simplified error scheme to hold, the
typical uncertainties in the relative mass of the ionizing cluster
with respect to the mass of the ionizing cluster are 20\%. The
uncertainties in the age of the underlying population are 25\% for ages
below $10^{9}$yr, and twice this amount for older ages. This is so because
of the flattening in the \bv-age relationship of the \citet{bruzual03} models.
The results are also presented in table \ref{res_1}.

\subsection{NICMOS NIR Photometric Masses.}

The final step in this crude description of the stellar populations found in the observed LCBGs
is made possible by the use of the NIR data. This information is now employed to check 
the likelihood of the derived stellar populations, by using the total luminosities
emerging from the line-emitting regions in the wavelength range that transforms into
the F160W wavelength interval measured by NICMOS to rederive the total
stellar masses contained in the line-emitting region, which is the area probed by the STIS spectra.
This is done by calculating the relevant $M/L_{X}$ ratios via the
$M/L_{B}$\footnote{The $L_{B}/L_{X}$ ratios are calculated for the (previously derived) 
model that best describes each particular object. The two families of models \#1 and \#2 are used.} 
ratios, making use of the programs presented in \citet{bruzual03}.
The statistical uncertainty that affects these NIR derived photometric masses
is a combination of two effects. The first effect is the intrinsic uncertainty
in the flux measurement, and the second effect comes from the scatter
in the $L_{B}/L_{X}$ ratio. This latter error source dominates over the first one, having
a scatter of 30\%. This number was derived by recalculating the $L_{B}/L_{X}$ number for
different BC03 spectra of the two metallicities considered for the underlying stellar
populations, with the typical ages predicted in the models previously constructed.

Total model masses, NIR photometric masses as well as the ratio between the two
are given in Table \ref{res_1}. It is seen in that table that the first set of models
produces NIR photometric masses that are in very well agreement (to within 25\%) with the 
masses derived using only the spectroscopic and optical photometry. The agreement for the
second set of models is somewhat worse (to within a factor of 2).
These results lead to think that the models \#1 are a better overall description.

\subsection{Mental model for the observed objects.}

Table \ref{res_1} lists the ages and masses of the stellar populations
that best represent the observed objects in the framework of the presented
models. This table also gathers the star formation rates and the
percentage of the dynamical mass that is accounted for by the stellar
populations used.

\begin{deluxetable}{lllllllll}
\tabletypesize{\scriptsize}
\rotate
\tablewidth{0pt}
\tablecaption{Model results. Upper half of the table gathers the first set of models, while models \#2 are given in lower half. \label{res_1}}
\tablehead{
\colhead{ID} & \colhead{1} & \colhead{1} & \colhead{2} & \colhead{2} & \colhead{3} & \colhead{3} & \colhead{4} & \colhead{4} 
} 
\startdata

Position.                           & inner & outer & inner & outer     & inner	   & outer &  inner &  outer  \\
Age of the ionizing cluster.(Myr)   & 3.1   & 3.2   & 4.8   & 4.9       & 3.2	   & 3.2   & 3.3    & 3.3     \\ 
\bv color of the ionizing cluster.  & -0.16$\pm$0.04   & -0.14$\pm$0.04   & -0.10$\pm$0.04   & -0.08$\pm$0.04       & -0.14$\pm$0.04	   & -0.14$\pm$0.04   & -0.15$\pm$0.04    & -0.15$\pm$0.04     \\ 
Age of the underlying population.(Myr)   & 500$\pm$100   & 1500$\pm$800  & 900$\pm$200   & 800$\pm$200      & 18$\pm$5     & 280$\pm$70  & 400$\pm$100    & 1300$\pm$600    \\
$\log M/M_{\odot}$.(Ionizing cluster). & 6.9   & 7.11  & 6.5   & 6.7       & 6.8      & 7.0   & 7.1    & 7.2     \\ 
Mass underlying/ionizing.           & 80$\pm$20    & 140$\pm$30   & 260$\pm$50   & 1340$\pm$300      & 9$\pm$2     & 100$\pm$20   & 80$\pm$20     & 190$\pm$40     \\

Stellar mass/Dynamical mass.	    & 0.20  & \dots & 0.60  & \dots     & 0.20     & \dots & 0.60   & \dots   \\
$\log M/M_{\odot}$ (STIS+WFPC2.)    & 9.4   & \dots & 9.9   & \dots     & 9.0      & \dots & 9.6    & \dots   \\
$\log M/M_{\odot}$ (NICMOS)         & 9.5   & \dots & 9.7   & \dots     & 9.1      & \dots & 9.7    & \dots   \\
SFR ($M_{\odot} \mathrm{yr}^{-1}$)  & 6.5$\pm$0.3   & \dots & 0.8$\pm$0.06   & \dots     & 4.7$\pm$1.0  & \dots & 6.8$\pm$1.2    & \dots   \\  
\hline

Position.                           & inner & outer & inner & outer     & inner	   & outer & inner  & outer   \\
Age of the ionizing cluster.(Myr)   & 2.9   & 3.2   & 4.9   & 5.1       & 2.9	   & 4.0   & 4.0    & 4.1     \\ 
\bv color of the ionizing cluster.  & -0.17$\pm$0.04   & -0.14$\pm$0.04   & -0.08$\pm$0.04   & -0.07$\pm$0.04       & -0.17$\pm$0.04	   & -0.12$\pm$0.04   & -0.13$\pm$0.04    & -0.13$\pm$0.04     \\ 
Age of the underlying population.   & 600$\pm$200   & 790$\pm$200   & 1000$\pm$500  & 1000$\pm$250       & 60$\pm$20       & 550$\pm$100   & 400$\pm$100    & 900$\pm$200     \\
$\log M/M_{\odot}$.(Ionizing cluster). & 6.9   & 7.11  & 6.5   & 6.7       & 6.8      & 7.0   & 7.1    & 7.2     \\ 
Mass underlying/ionizing.           & 70$\pm$10    & 90$\pm$20     & 500$\pm$100    & 800$\pm$200        & 28$\pm$6        & 220$\pm$40    & 90$\pm$20      & 250$\pm$50      \\
Stellar mass/Dynamical mass.	    & 0.05  & \dots & 0.40  & \dots     & 0.40     & \dots & 0.70   & \dots   \\
$\log M/M_{\odot}$ (STIS+WFPC2.)    & 9.2   & \dots & 9.7   & \dots     & 9.4      & \dots & 9.7    & \dots   \\
$\log M/M_{\odot}$ (NICMOS)         & 9.6   & \dots & 9.9   & \dots     & 9.5      & \dots & 10.0   & \dots   \\
SFR ($M_{\odot} \mathrm{yr}^{-1}$)  & 6.5$\pm$0.3   & \dots & 0.8$\pm$0.06   & \dots     & 4.7$\pm$1.0  & \dots & 6.8$\pm$1.2    & \dots   \\  

\enddata


\tablecomments{Ages and masses of the ionizing population are lower limits to their true values. The same
can be said about the total masses. The statistical uncertainties of the optically derived masses are 
around 10\%, although the systematic errors are probably larger.
The statistical error affecting the NIR derived masses are around 30\%.}

\end{deluxetable}

The star formation rate (SFR) was calculated from the H$\alpha$ luminosity as in
\citet{ken94}, valid for T$_{e}=10^{4}K$ and case B recombination (all
the ionizing photons are processed by the nebular gas).
The results are given in Table \ref{res_1}.

\begin{equation}
SFR(M_{\odot}yr^{-1})=7.9\times 10^{-42} L_{H\alpha} (erg\ s^{-1})
\end{equation}

Although the preceeding SFR estimate was based on spatially averaged SFR for disk galaxies, we believe 
it can be applied for the current sample of starburst galaxies. The $H\alpha$ 
derived SFR only probe the most recent star formation episodes of any given 
galaxy because the $H\alpha$ line becomes very faint soon after the more massive stars, born in the more 
massive starbursts, die out, and this is true for all normal galaxies. For this reason, it is possible 
to use the \citet{ken94} expression for the global SFR of actively star forming galaxies, too.
In fact, the SFR $H\alpha$ estimate has been calibrated against many other instantaneous star 
formation tracers \citep{rosa02}.
Differences between the \citet{ken94} $H\alpha$-derived SFR and the real SFR of the studied galaxies are 
bound to arise mostly from other issues, such as differences in the IMFs used.

The calculated star formation rates are in the range from 0.8 to 6.8
$M_{\odot}$yr$^{-1}$, indicating that the star-forming episode is very
strong (30-Dor SFR is 0.1 $M_{\odot}$yr$^{-1}$). It has, however, to be said
that the line-emitting regions typically span around 3 kpc, and the 30-Dor 
line-emitting region is a only few hundred parsecs. The star formation rates
per unit area are therefore of the same order of magnitude.

Tables \ref{res_1} shows that the ionizing populations
are very young, less than 5Myr in almost all cases, although the
average age is slightly over 3Myr. The ionizing clusters are then
largely unevolved. The resulting ages in both model sets indicate that
these galaxies are being observed at the peak (or a little bit past
of) of their line luminosities. It is also the case that the ages of
ionizing populations are very similar for the inner and outer regions
of each galaxy.

The inferred underlying populations are very different amongst the
studied galaxies. There are systems with fairly young non-ionizing
populations (around 300Myr old), but there are LCBGs whose underlying
populations are very old (older than 1.0Gyr). In the case of the inner
region of SA57-10601, the estimated stellar population is compatible
with a single ionizing stellar population.  The modelled age of the older
stellar population is usually larger for the outer regions than for
the inner regions of each object. However, given the uncertainties
in the ages estimates for the underlying populations (between 25\% and 50\%), it
is not possible to extract much information from these ages alone.
The average age of the underlying stellar populations found are 700Myr
for both models \#1 and \#2.

The derived stellar masses account for 20\% to 60\% (the average value
being 40\%) of the virial masses \emph{within $R_{e}$}, according to the
first set of models. In the case of models 2, the stellar mass accounts
for 4\%--70\% (the average value is again 40\%) of the derived virial
masses, where the virial masses were determined as in
\citet{hoyos04}, assuming the measured gas-phase velocity dispersion
equals the star's velocity dispersion and that LCBGs are in dynamical
equilibrium. These are the virial masses within $R_{e}$. 
Since the mass estimates yielded by the models used here are only lower limits
for the reasons outlined 
above\footnote{We remind the reader that the masses of the underlying populations
are scaled with respect to the mass of the ionizing population. This is calculated
via $W^{0}_{\beta}$, which is in turn a lower limit to the real equivalent width. Other factors
such as photon escape, dust and enhanced photon production have an impact, too.}, there is not much
room left for the presence of dark matter in the innermost regions of
these objects. The dark matter would, in any case, be spread out in a
vast halo around the objects by its very nature. The presence or not of significant amounts
of dark matter in the optical cores of these systems is important. If the 
conjecture introduced here that these systems are not dominated by dark matter in their very centers
turns out to be true, it would imply that these systems could not possibly evolve
into today's population of Low Surface Brightness objects.
A very simple and crude estimate based in the \citet{mamonlokas05} models and
the \citet{nav04} simulations indicate that the mass of dark matter
within a 1.5 kpc sphere\footnote{Similar in size to $R_{e}$.} is 
$\sim2 \times 10^{8}\pm \mathrm{40\%}$, assuming that the total mass of the dark
matter halo is ten times the measured kinematic mass. This contribution is much less 
than the kinematical masses derived from velocity
dispersions.  In order to determine the dark matter content and
density profile of these sources with greater accuracy, measurements
of the line of sight velocity distribution out to several (around
5--9) effective radii would be required. Such data are of course not
available for the distant objects studied here.

It might be tempting to think that the remaining 60\% of the dynamical mass not
accounted for by the stellar component might be hinting the existence
of a very old stellar population, too faint and red to leave a trace in the \bv
color used here but as massive as the already detected underlying
stellar population. However, this hypothetical phantom population can not be older
than $\sim5 \mathrm{Gyr}$, simply because of look-back time reasons,
implying that such very old and massive population can not possibly
exist. Such population would still be shining in the \bv
color\footnote{Assuming these stars occupy a volume similar to the
other populations, i.e., the surface brightness of this population is
not lowered for other reasons} because the turn-off point of this
population would be around solar-type stars or slightly earlier.
More strongly, the good agreement between the optically derived masses
and the NIR derived masses sets very strict limits to the luminosity
of this possible population, virtually forbidding it for the case
of models \#1 and for some of the models \#2 as well.
Finally, the presence of significant amounts of neutral or molecular
gas in the central regions of the sources studied can be safely discarded. It is very likely
that the extreme starburst taking place at the cores of these objects is well capable of
fully ionizing the ISM in the innermost zones of these galaxies.

It is therefore concluded that the stellar inventory found using the simple
models devised here can not be very far from reality and that the
disagreement between the stellar masses and the dynamical masses
arises from the fact that the mass calculations for the stellar
component are forcibly lower limits.

The models constructed here are similar to the two-bursts models
presented in \citet{guz98}. In the models presented here, the ionizing
cluster is less massive than the ionizing clusters used in that
work. In addition, the underlying stellar populations used in
\citet{guz98} are much older than the ones derived using the models
presented here.
The relative masses of the ionizing populations
considered in \citet{guz98} went from 1\% to 10\%, while the results
presented in this paper lead towards much lower relative masses
for the newer stellar generation, between 0.1\% and 3\%.
Another very important difference is that the underlying stellar
population used in \citet{guz98} was set to be 3Gyr old at the time of the ionizing
generation birth, whereas in the current work this age is found to be much shorter, around 1Gyr.

Finally, this section is closed with a brief summary of how the models are
built. This procedure is carried out for each region (a) and (b), and assumed metallicity
of each object.
See \citet{diaz00} for a complete account on how to mix several single 
stellar populations to match the observed properties of star
forming regions.

\begin{itemize}

\item{The first step is to derive the H$\beta$ ``effective'' equivalent width.
This is done via the ionization ratio [\ion{O}{3}]/[\ion{O}{2}]
and the relationship given above.}

\item{This ``effective'' equivalent is assumed to be equal to the ``intrinsic'' value
and used to determine the age and \bv color of the ionizing stellar population. In this step
the error budget is dominated by the systematic error of assuming that both quantities
are equal.}

\item{Using the measured $H\beta$ luminosity, the mass of the ionizing population is 
estimated. This estimate can only be a lower limit to the real value because some unprocessed 
photons might be escaping from the nebula, and dust globules will be present. This also
translates into the derived masses for the underlying stellar populations.}

\item{Using the \emph{emission-line free} \bv color and the \emph{observed} $H\beta$ equivalent width, the
age and relative mass of the noionizing population are derived. The errors in these quantities are dominated by the
uncertainties in the required color and required specific intensity at 4860\AA of this population.}

\item{The NIR observations are then used to check if the models created to describe the optical
properties of the observed galaxies also describe their NIR properties. It is the case that
the agreement is excellent for the metal-rich models, and somewhat worse for the metal-poor ones.}

\end{itemize}

\section{Comparison with Local Galaxies: Bright, Local H\textsc{II} Galaxies as Nearby LCBGs and the LCBG-Spheroidal Connection.}
\label{compa}

It is interesting to compare the evolution of the models described
here with present-day galaxy samples. In this section, several
predicted observables are compared to the properties found in local
systems. In particular, the issue of whether or not the stellar
populations predicted by the presented models are similar to the
underlying stellar populations of local \mbox{H\textsc{II}} galaxies
or in the less luminous Blue Compact Dwarf Galaxies is addressed. This
section also deals with the passive evolution of LCBGs, and their
connection with the local population of dwarf elliptical systems.

The \mbox{H\textsc{II}} galaxy sample used here was taken from the
\citet{vri_telles97} sample.  This sample comprises 15 galaxies, which
were selected from the Spectrophotometric Catalogue of
\mbox{H\textsc{II}} galaxies \citep[SCHG]{T91}. Their equivalent
widths are between 26\AA \ and 170 \AA \,
and their M$_{B}$ ranges between -14.5 and -21.5, although with a preference
for the brighter values. The measured velocity
dispersions range from 16 km s$^{-1}$ to 50 km s$^{-1}$.  The galaxies
also exhibit a wide morphological variety.  This sample was chosen
since \mbox{H\textsc{II}} galaxies also show conspicuous and
galaxy-wide emission line spectra. They are quite similar to the LCBGs
presented here in this respect, as hinted by \citet{koo94},
\citet{guz97} and \citet{hoyos04}. Their small radii and velocity
dispersions indicate that they ought to be as massive as distant
LCBGs. In addition, they are also known to posses violent star forming
systems, just like intermediate-\textit{z} LCBGs.  It is then natural
to wonder if bright, local \mbox{H\textsc{II}} galaxies are nearby
copies of distant LCBGs or if they are different from the distant
sources in some respect.

The BCDG control sample is the one found in the
\citet{doublier397,doublier299} works.  This sample consists on 44
galaxies, which were selected from the Second Byurakan Survey
\citep{mark67} and other lists. The \citet{caon05} BCDG sample was
also included. It adds 7 objects.  Their luminosity profiles are
typically $r^{1/4}$ profiles, although there is a sizeable
fraction of purely exponential profiles and mixtures of both
profiles. Their star forming regions are clearly detected in $B-R$
color maps and show an age spread. The presence of very old red giant
stars indicates that these systems began to form their stars several
Gyr ago, at the time in which the intermediate redshift LCBGs were
observed.  This sample was chosen since BCDGs could be considered to
be the low-luminosity end of \mbox{H\textsc{II}} galaxies in many
respects. Therefore, they can be used to make educated guesses on the
properties of other populations of faint intermediate redshift objects
whose luminosities are not high enough to be considered LCBGs and are
therefore very difficult to study observationally, given their
distances.

It is also possible to investigate the \emph{passive} evolution of
these systems in the framework of the presented models. This scenario
of a passive evolution assumes that no further episodes of star
formation takes place from their look-back times to the present
day. Given the high intensity of the observed starburst, it is not
unlikely that the interstellar medium will be cast out of the galaxy,
preventing further starbursts from happening. However, it has to be
borne in mind that the look-back times of the LCBG sample used here
are $\sim 5\times 10^{9}$yr. This means there is a huge amount of time
for they to undergo other star formation episodes, mergers or other
phenomena that would, of course, not be represented by the simple
passive evolution scenario used.  In particular, the evolution of the
intermediate redshift LCBGs studied here into today's dwarf elliptical
systems is studied.  The dE/Sph sample used for comparison purposes is
that of \citet{guz_tesis94}. We also use four other well-known dwarf
elliptical galaxies from the local Universe. The observed LCBGs
galaxies are also compared with The Large Magallanic Cloud.

It is important to note that the available colors and magnitudes for
the comparison samples are \emph{integrated} data while the models for
the LCBG sample have been constructed from the line ratios and colors
of the line emitting region of each object. The
model results for the inner and outer regions of each galaxy have been
therefore mixed together and scaled using the appropriate weights to
derive the properties of the whole sources. The main assumption is that
the \emph{underlying} stellar population of the central kpc of these objects
is qualitatively similar to the stellar population found in the
outskirts\footnote{Here, the word ``outskirts'' means ``outside the
line-emitting region.''} of each galaxy.
The absolute luminosities were further corrected by multiplying them
by a factor of two, to account for the luminosity of the stars found
in the outskirts of the studied
objects\footnote{This factor is needed because the STIS spectra only samples
within $R_{e}$}. This way, the observed
galaxies can be represented by a single point each in color-color or
color-magnitude diagrams.

\begin{figure}
\plottwo{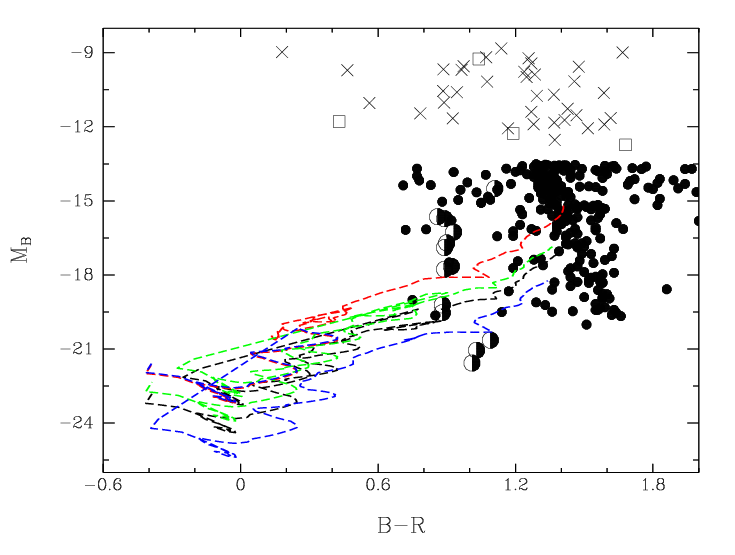}{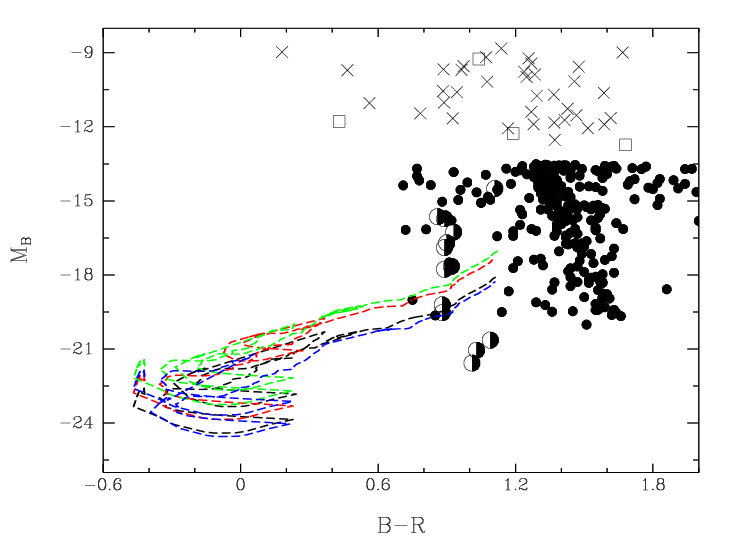}
\caption{Model evolution in the $(B-R)$--$M_{B}$ plane. The green
line represents the evolution of the SA57-7042 models, the blue line
represents SA57-5482, the red line shows the evolution of SA57-10601
and the black line is H1-13088. The black dots represent the dwarf
elliptical systems from \citet{guz_tesis94}. Half filled dots are
local \ion{H}{2} galaxies from \citet{vri_telles97}. Crosses and white
squares represent the local BCDG samples from
\citet{doublier397,doublier299,caon05} respectively. The left panel
represents the metal-rich models whereas the right panel depict models
\#2.\label{mb.br_main}}
\end{figure}

Figure \ref{mb.br_main} presents the evolution of the stellar
populations described by the models used here from their predicted
creation time to the present day in the $M_{B}$--$B-R$ plane for the
two sets of models used. The $B-R$ color information is an output
from the \citet{bruzual03} programs. The $B$ magnitude is 
the standard Johnson magnitude while the $R$ magnitude is the
Cousin's $R$ magnitude.
Each intermediate redshift LCBG is
represented by a colored line resulting from appropriately combining
and scaling the models for the inner and outer zones of the
line-emitting region, as stated above. This is also true for the rest
of the diagrams.  The Sph sample is represented by simple filled dots
and the local \ion{H}{2} galaxies are represented by half-filled
circles. The BCDGs from the \citet{doublier397,doublier299} samples
are depicted as crosses. The empty squares represent the
\citet{caon05} galaxies.

The $B-R$ colors plotted in figure \ref{mb.br_main} for the \ion{H}{2} galaxies
and BCDGs are not the $B-R$ colors of the \emph{whole} galaxies,
rather, they are the $B-R$ color of the \emph{underlying} stellar
component, excluding the starburst.
This $B-R$ color has been estimated as the terminal color in
$B-R$ color profiles, or from the $V-R$ colors of the extensions,
assuming \bv=0.65. The $B-R$ color of the dE sample is the color of the 
entire galaxy, since there is no ionizing population in these objects.
Therefore, it is possible to compare both $B-R$ color estimates in the sense
that, if the $B-R$ color of the wings of an \ion{H}{2} galaxy or BCDG
is similar to the color of a Sph system, it would be possible
to think that the starbursting \ion{H}{2} galaxy or BCDG is a Sph system
experiencing one further starburst. Figure \ref{mb.br_main} shows that the 
underlying stellar population of \ion{H}{2} galaxies and most 
BCDGs are younger than the average $B-R$ color of Sph systems, implying that
there are have to be other differences between these galaxy types.
In the case of the \ion{H}{2} galaxies and BCDGs, the plotted values of 
$M_{B}$ are the blue absolute magnitudes of the \emph{stellar} component
only. These values were derived from the \emph{global} $M_{B}$ using the appropriate 
available data in the references used, and were corrected for galactic extinction according
to \citet{bh82}. The values of $M_{B}$ for the LCBG models were
normalized to the starburst's luminosities, which are common to both
the metal rich and metal poor models. The distance modulus used to derive the
blue absolute magnitudes of the \citet{guz_tesis94} sample from their reported apparent magnitudes
is $35.0\pm0.3$, which is the average value of the
\citet{alexterlevich01} sample\footnote{Both studies dealt with the 
Coma Cluster, which is located at 100Mpc.}. These 
transformations ensure that, when comparing all these samples with the LCBG models at $z=0$,
colors and magnitudes are free from the contribution of emission lines and the subsequent comparisons
are between \emph{purely stellar} properties only.

It can be seen in figure \ref{mb.br_main} that the first set of models
ends its evolution much closer to the bulk of the Sph systems than the
second set of model does.
In fact, should the metallicity of LCBGs be on the lower end of the range
allowed by their line ratios, the simple passive path assumed here
could not transform them into Sph galaxies.
To illustrate this point, a third, 10 Gyr old stellar population
was added to models \#2. The mass of this population was
arbitrarly set to be 50\% of the dynamical mass. Two different metal contents
were tried for this primeval population, $Z=0.0004$ and $Z=0.0001$ since 
higher metallicities are not justifiable in the framework of these metal-poor 
models. This population turned out to be unable to change the final $B-R$
color to a great extent, affecting only in the scale of 0.05
magnitudes. This possibility was not explored further, then.
It is concluded that, if these intermediate redshift LCBGs
are to become Sph galaxies following a simple passive evolution, the first
set of models is likely to be a better approximation to their stellar
inventory. In order to reach to more robust conclusions on the metallicity of LCBGs
and their descendants, it would be needed to compare the metallicity of LCBGs
with that of Sph galaxies. However, the metal content of Sph galaxies
is very difficult to determine spectroscopically because of their low surface brightnesses, and 
only broad ranges are reported in the literature, mainly based in photometric 
measurements (see, e.g \citet{rakos03,jerjen04}). In the former work, the [Fe/H] metallicity 
of dE galaxies is found to lie in the $1/100$ to $1.0$ solar interval, whilst in the latter work
the iron content of dE galaxies is found to range from $1/25$ to $1/3$ of the solar value.
These broad ranges prevent us from using these values to discriminate between the
presented models. If solar proportions are assumed, the previous iron content ranges roughly 
translate into $12+\log \mathrm{O/H}$ ranging from 7.0 to 8.5. This range
is unfortunately in agreement with both the metallicities of models \#1 and \#2.

It is also seen in figure \ref{mb.br_main} that, even though BCDGs are
less luminous than Sph systems or \ion{H}{2} galaxies, their $B-R$
colors are similar although somewhat bluer on average. BCDGs are as
luminous as the brightest globular clusters, and their stellar
populations are also vey old, like those of Sph systems.
If they are the descendants of a hypothetical class of low-luminosity LCBG-like
galaxies in the burst phase, the blue absolute magnitude of these objects
should be around -16.0. Such objects would be very hard to observe and they
would very similar to dwarf irregular systems like NGC1569 or NGC1705.

Figure \ref{mb.br_main} shows that the underlying stellar populations
of the local \ion{H}{2} galaxies are compatible with the evolution of
LCBGs in the sense that their $B-R$ colors were the same at some point
in the past around the time when the observed starbursts took place.
It is then likely that the underlying stellar populations found in the 
presented LCBGs are much younger than the underlying stellar generations found
in local \mbox{H\textsc{II}} galaxies. The underlying
stellar population of local \ion{H}{2} galaxies is found to be older
that 1 Gyr, in the work of \citet{westera04}. The older stellar generations
derived in the simple models presented gere resemble a somewhat aged intermediate
age population, again in the nomenclature given in \citet{westera04}.

\begin{figure}
\plottwo{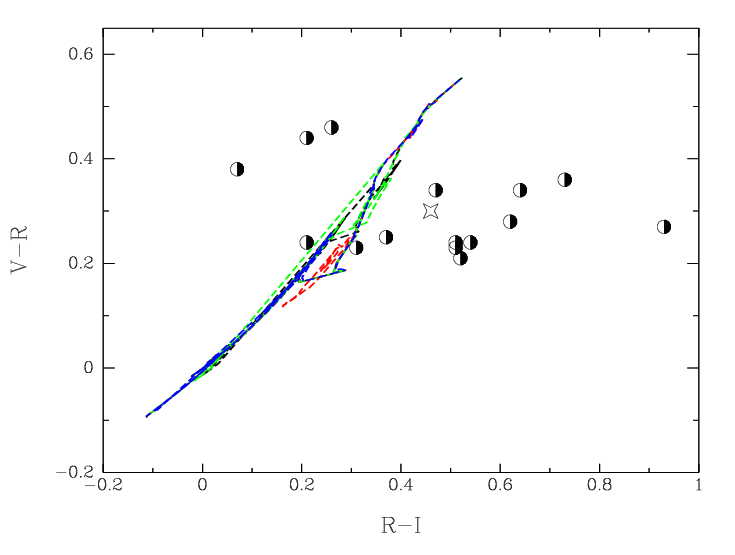}{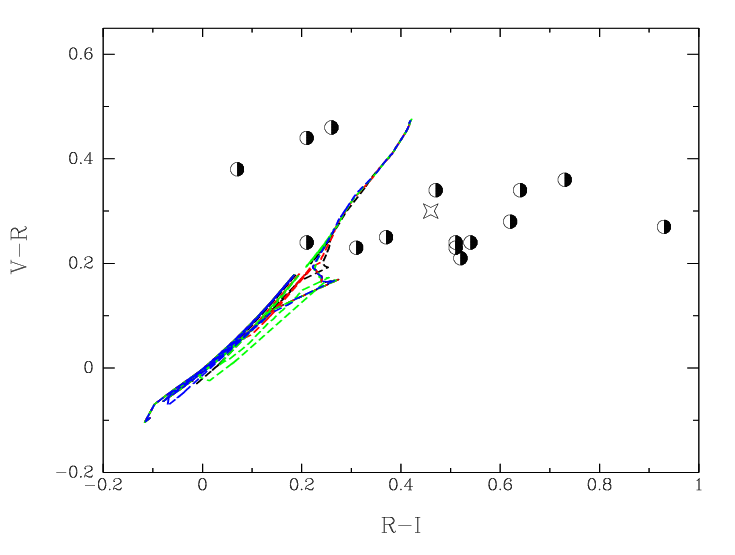}
\caption{Both panels represent the evolution in the $V-R$--$R-I$ plane
of the models used here to describe the four observed LCBGs. The left
panel represents the first set of models, while the right one shows
the metal-poor models.  The local data are taken from the
\mbox{H\textsc{II}} galaxy sample from \citet{vri_telles97}.  The
square star represents the average position of this \ion{H}{2} galaxy
sample.  Other symbols as in figure \ref{mb.br_main}.\label{comp3}}
\end{figure}

In order to compare to a greater detail the stellar populations of
LCBGs and \ion{H}{2} galaxies, the evolution in the ($V-R$)--($R-I$)
plane of the models devised here is presented in figure \ref{comp3},
together with the location of the underlying stellar populations of
the 15 bright, high equivalent width \ion{H}{2} galaxies from \citet{vri_telles97}.
These latter objects are again represented by half-filled dots. The $V-R$
colors are in a very narrow range around 0.3, the $R-I$ color has a
much greater range of variability, from 0 to 1. This figure also shows
that the $V-R$ color of the underlying population found in local
\ion{H}{2} galaxies is indeed attainable by LCBGs at some
point during their evolution\footnote{In particular, according to
models \#1, the underlying populations of LCBGs presented such $V-R$
colors 4.9 Gyr ago, when the observed starbursts took place.
In the case of the metal poor models, this happens 4.5Gyr ago, just
after the observed starbursts occured. The first set of models reaches 
the average $R-I$ of the local sample 2.3Gyr ago, while models \#2 
will attain such $R-I$ color in the future.}. 
The underlying stellar populations of local
\ion{H}{2} galaxies are then found to emit more power in the I band than
intermediate redshift LCBGs of similar $V-R$ colors.  This might be
caused by the fact that intermediate redshift LCBGs probably harbor
lower \emph{relative} numbers of red giant stars than the redder
\ion{H}{2} galaxies since they have had much less time to
evolve since the onset of the red giant phase of the evolution of
their dominant stellar populations. Such very luminous red giant
stars, existing in the extensions and wings of the local H\textsc{II}
galaxies, would be responsible for a sizeable fraction of the I band luminosity of
local \ion{H}{2} galaxies.

These suggestions can be interpreted in the light of the picture for 
\ion{H}{2} galaxies presented in \citet{westera04} in which
\ion{H}{2} galaxies are a mixture of three populations. The first of them is the youngest
population responsible for the observed emission lines, the second population
is an intermediate age generation, older than 50Myr but younger than a few hundred Myr and the third
population is the oldest one, older than 1Gyr. The single underlying stellar
population used here can be thought to represent a mixture of the second
and third populations of the \citet{westera04} model. As such, it can only accurately
reproduce the colors of the real mixture it represents in a limited wavelength
interval. Figure \ref{comp3} shows that the model fails in the I band. It is then natural to think
that the real stellar mixture in \ion{H}{2} galaxies is relatively richer in very 
old, I emitting stars, namely red giants.
This very old and important stellar component has already been shown to exist 
in local \ion{H}{2} galaxies. In addition to this, the work presented in \citet{raimannXX} showed
a detailed statistical analysis of the stellar populations
found in a large sample of emission-line galaxies. It was found that
a significant fraction of the \emph{stellar} mass built in these systems is at least 500Myr old.
Present day descendants of LCBGs will then have very old stellar populations, similar
to the oldest stellar populations described in \citet{westera04}. Should these systems
undergo another star formation episode they would be very similar to modern \ion{H}{2} galaxies.
More modelling and observational work is clearly needed in order to ascertain this.

\begin{figure}
\plottwo{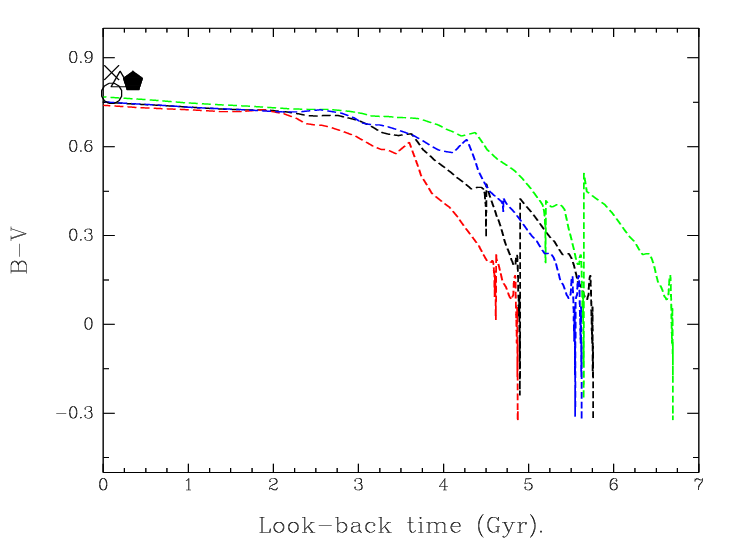}{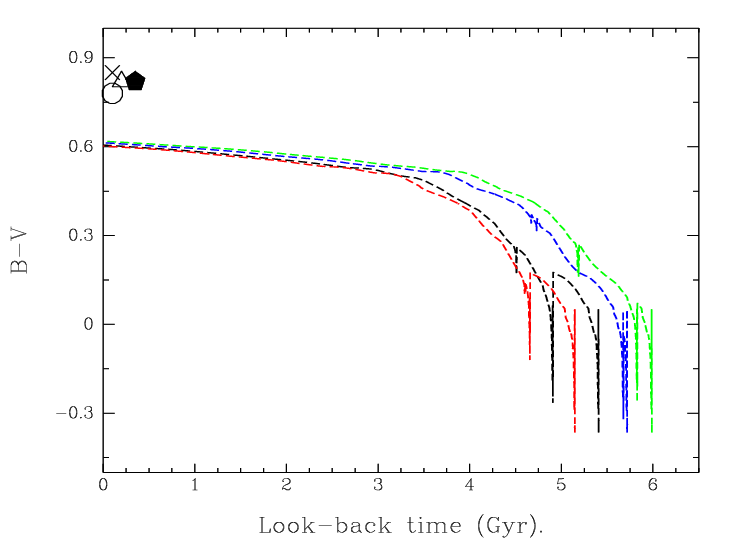}
\caption{Temporal evolution of the stellar \bv color with look-back
time. The left panel represents the first set of models, while the
right one shows the metal-poor ones. Individual symbols represent the
\bv color of several well known local spheroidal systems. The black
pentagon is NGC205, the white triangle is IC3393, the cross is NGC3605
and the white circle is NGC147. Other symbols as in figure
\ref{mb.br_main}.\label{comp.bv.lb}}
\end{figure}

Figure \ref{comp.bv.lb} presents other
aspects of the evolution of the constructed models.  It 
shows the evolution of the \emph{stellar} \bv color
as a function of look-back time. The symbols at $t\sim 0$ represent
four well known spheroidal systems, as indicated in the figure. The
metal rich models end their evolution near the position occuppied by
the dE/Sph sources. It can also be seen that the second set of models
does not reach the very red colors of the presented Sph
galaxies. Moreover, these metal poor models can only reach \bv$=0.7$
in the distant future. Such colors are not in agreement with the
colors of the handful of Sph galaxies presented for comparison.

\begin{figure}
\plottwo{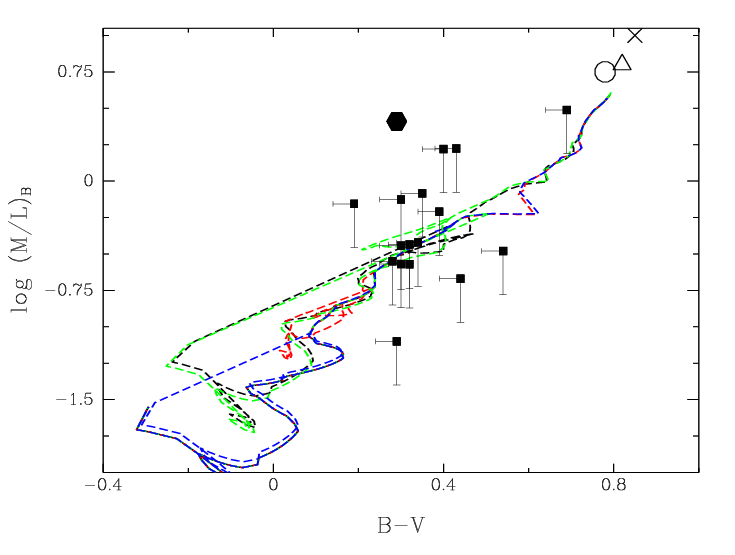}{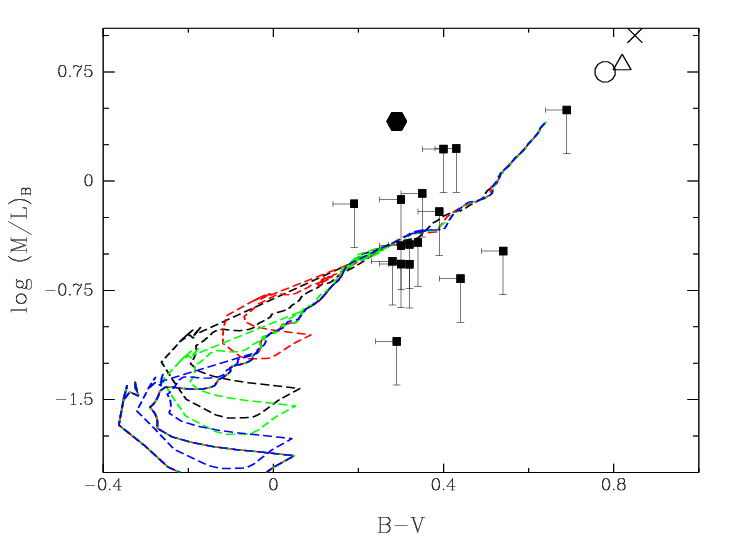}
\caption{Model evolution in the \bv--$(M/L_{B})$ plane.As before, the
left panel represents the first set of models, while the other panel
represents the metal-poor models. The position of the galaxies with
$M/L$ ratios taken from \citet{guz98} is given as small squares, and
their error bars are explained in the text.  Individual symbols
represent the position of several well known local spheroidal
systems. The Large Magallanic Cloud is included, too.  The white
triangle is IC3393, the big cross is NGC3605, the white circle is
NGC147 and the black hexagon is the LMC.  Other symbols as in figure
\ref{mb.br_main}.\label{comp.bv.mle}}
\end{figure}

In figure \ref{comp.bv.mle}, the evolution of the presented models in the
$\log (M/L_{B})$--(\bv) diagram is shown. The positions of the LMC and
individual dwarf elliptical systems is shown with specific symbols for
each object. The position of the LCBGs observed with the WF/PC-2 in
\citet{guz98} is depicted using small squares with asymmetric error
bars.  The errors in $M/L$ arise from the fact that the masses used in
\citet{guz98} were not stellar masses but dynamical ones. The
presented error bars allow for a factor of two difference in both
numbers.  The errors in the \bv color come from the contribution of
the emission lines to the spectrum. This contribution has been assumed
to be 0.05 magnitudes towards bluer colors for all objects.  The \bv
color of the LMC has been corrected for the contribution of the
line-emitting region and starburst 30-Dor and extinction by
subtracting 0.15 to the color.

It is again seen that models \#1 end their evolution much closer 
to the Sph galaxies, although neither
model falls neatly within the region defined by the Sph galaxies
shown.  The fact that the mass-to-light ratios of the first set of
models are more similar to those of the Sph systems than those of
models \#2 allows us to think that the rest of the other colors of dE
will be more similar to the colors predicted by models \#1.  It is
also the case that the stellar populations of the presented models
have a mass-to-light ratio lower to that of the LMC by the time their
\bv color is similar. This suggests that the star formation has
proceeded for a longer period of time in the LMC than in LCBGs, which
is again consistent with LCBGs being observed at very high look-back
times.  The \citet{guz98} objects cluster around the model tracks. In
fact, six objects lie directly on top of the models themselves. This
is reassuring in the sense that the bulk of the stellar inventory in
the \citet{guz98} sources is probably well represented by the simple
models presented here.  However, it is somewhat unfortunate that the
\citet{guz98} points fall both above and below the model lines because
this makes it impossible to further resolve the underlying stellar
population.  If, for instance, the \citet{guz98} points were all above
the model tracks around the LMC, it would have been possible to argue
that the real underlying stellar population should be older than the
age predicted by the models. This fact, however unfortunate, was to be
expected since the \bv color is not very informative in the age range
studied. The use of redder colors would be needed in order to gain
this drop of information on the underlying stellar population. 

It is also interesting to compare figure \ref{comp.bv.mle} with 
\citet[][Fig 3. of]{guz98}. It is seen that the path of the models built in this
work in the $\log (M/L_{B})$--(\bv) plane is very similar to the path
of the \citet{guz98} models, implying that the differences in the age
of the underlying stellar generation and relative mass of the ionizing
population have little impact on the \bv color and Mass-to-Light
ratio, at least within these limits. Other observables will surely be
more affected.

\begin{figure}
\plottwo{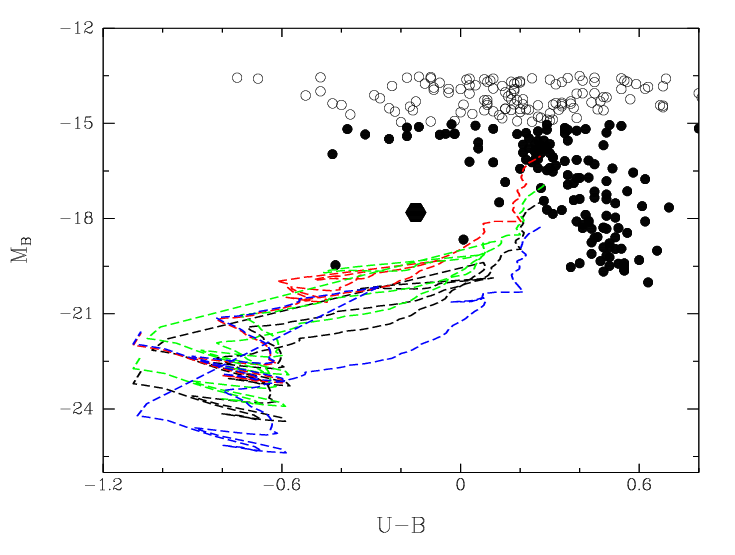}{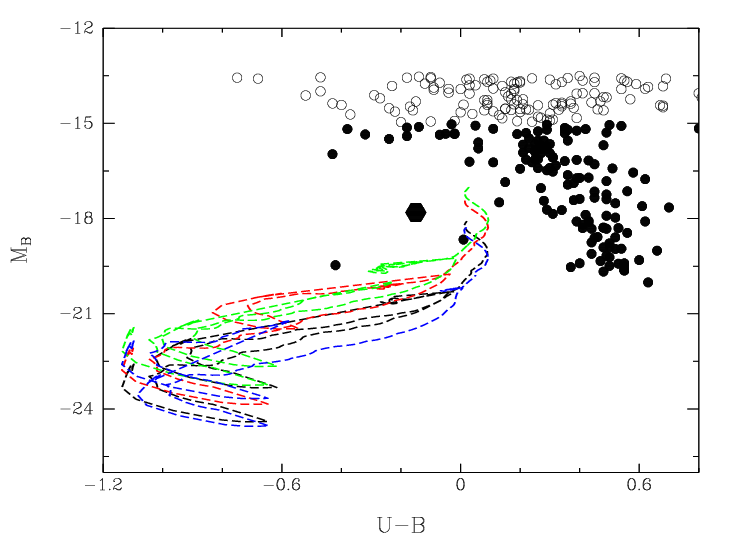}
\caption{Model evolution in the ($\ub$)--$M_{B}$ plane. The Large
Magallanic Cloud is included as a black hexagon. 
Both open and closed circles represent the Sph galaxies from \cite{guz_tesis94}. Closed symbols
represent objects brighter than $M_{B}=-15.0$. Open symbols are the dimmer objects.
\label{comp.Mb.ub}}
\end{figure}

Figure \ref{comp.Mb.ub} is very similar to figure \ref{mb.br_main}. It
shows how the presented models evolve in the $M_{B}$--(\ub) plane.
This time, only the position of the LMC and of the sample of
spheroidal systems taken from \citet{guz_tesis94} is included. The
\ion{H}{2} galaxies and BCDGs are not included because there are not U
band data for these galaxies, and estimating it from V band data would
be affected by large uncertainties. The contribution to the \ub color
from the 30-Doradus cluster and extinction is assumed to be 0.1. It is
again seen that the final stage of the first set of models is much
more similar to the observed properties of the Sph
galaxies. Furthermore, all the models converge towards a point around
$\ub=0.3$, $M_{B}=-17.0$.  This suggests that any Sph descendant of
the distant LCBGs should be the very red and luminous.  This latter
conclusion is weaker in figure \ref{mb.br_main}.  However, since the
model used is very simple, it is also likely that
the aforementioned convergence towards this subset of the Sph
galaxies merely reflects a weakness of the models. Indeed, should any
major star forming episode happen in these objects, the final position
of the evolved LCBGs would be different. It is also interesting to
note that the LCBG models run very near the LMC at some point in the
past, some time after the observed starbursts took place.  A larger
sample of LCBGs with more data would be needed to study the reality of
the convergence effect mentioned above.

\begin{figure}
\plottwo{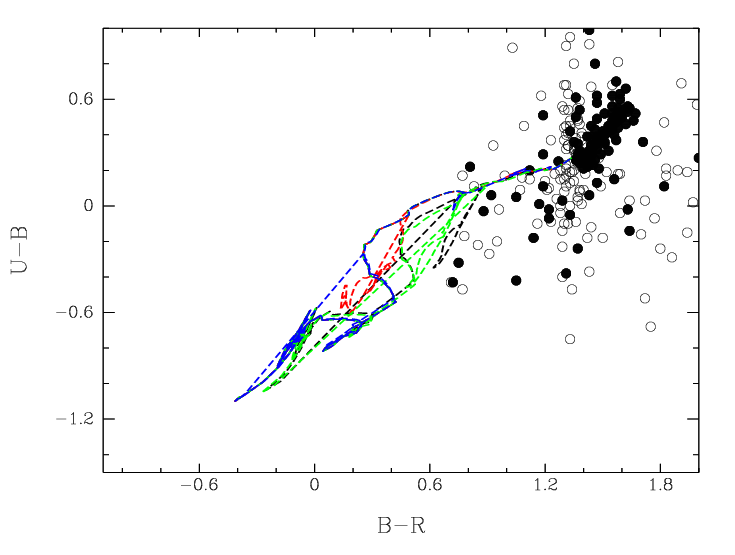}{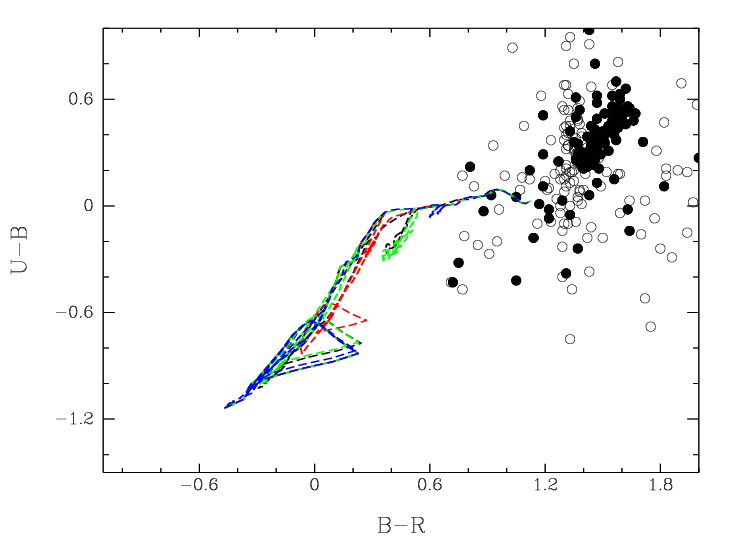}
\caption{Evolution in the $(U-B)$--$(B-R)$ plane of the stellar
populations that best represent the observed LCBGs. Both open and closed
circles represent the Sph galaxies from \cite{guz_tesis94}. Closed symbols
represent objects brighter than $M_{B}=-15.0$. Open symbols
are the dimmer objects. Panel layout as in figure \ref{mb.br_main}.\label{comp.ub.br}}
\end{figure}

Finally, figure \ref{comp.ub.br} shows the
evolution of the stellar populations predicted by our models in the
(\ub)--($B-R$) plane. It is again seen that the first set of models ends its
evolution much closer to the Sph galaxies than the metal poor
ones. Moreover, the models converge towards the bulk of the dwarf
elliptical galaxies, and this conclusion also holds when only the
bright Sph systems are considered, strengthening the idea that the Sph
descendents of LCBGs are the more luminous ones.

\section{Summary and Conclusions.} \label{summ}

We have used spatially resolved STIS long-slit spectroscopy together
with optical and NIR imaging to investigate the stellar populations found in the
the line-emitting region of four LCBGs at intermediate \textit{z}.
This is one of the first times this has been done for distant, compact
galaxies.
The starbursting episodes are found to be less than 3.5 Myr old in most cases, with typical
masses in the range from $5\times 10^{6}$--$12 \times 10^{6}$ solar masses. The mass contribution
of the newly formed stars to the total mass of the stellar component goes from 0.1\% to 4\%, with an average
value of 1\%. The ages of the older stellar generations found in these systems are in the
20 Myr to 1400 Myr, with an average value of 700 Myr.
The NIR observations were used to check for consistency
the optically derived stellar populations. It turns out that the agreement
is generally very good, implying that the real stellar content
is bound to be very similar to the two population mixture
used here.
The line-emitting region spans $\sim2$ kpc in size and is
found to be very concentrated in the observed galaxies, although it is
not always in the center of the optical galaxies. The metal content of
the ionized gas phase can not be readily determined with the data at
hand, although the nitrogen measurements of very similar objects
suggest slightly subsolar abundances. The NICMOS data 
allow to regard the metal rich models as the favored ones, too.

Although the available data are very limited, we have shown evidence for 
a very important underlying stellar population even in the 
central region of these sources, where
the starburst completely dominates the observed optical emission. This
population is somewhat evolved and accounts for most of the mass of
these systems, despite the high look-back times probed.  It can be
said that the presented star formation histories are similar to some
extent to the star formation histories of \mbox{H\textsc{II}} galaxies
described in \cite{westera04}. Both \mbox{H\textsc{II}} galaxies and
LCBGs are therefore age-composite systems. They both have an old
stellar population, together with a young stellar generation.
It has also been found that, should LCBGs finally evolve into dwarf
ellipical galaxies, their metal content must be in the higher end of
the interval allowed by their observed line ratios.
Unfortunately, given the broad range of the metallicity determinations
of Sph systems, this information can not be used to help rule out any one of
the presented models. The NIR data point towards the higher metallicity 
ones, though
 In addition, it's been concluded that the Sph descendants
 of LCBGs can only be the more luminous and red of such systems.
The comparison of the LCBG models with the underlying population of
luminous and local \mbox{H\textsc{II}} galaxies indicates that the
star formation histories of the observed LCBGs and local
\ion{H}{2} galaxies are similar. It has to be kept in mind,
however, that \ion{H}{2} galaxies probably have a larger
relative number of red giant stars, as hinted by their enhanced power
in the I band. This can be simply interpreted by the fact that they
are, arguably, 5 Gyr older than the more distant LCBGs.
The observed differences in the redder colors probably arise from
differences in the relative numbers of red giant stars. This suggests
that the recent star formation histories of LCBGs and \ion{H}{2}
galaxies are similar up to some point. From that point backwards, the
star formation histories probably differ. It has also been pointed out
the possible existence of dwarf irregular systems in the intermediate
redshift range.

\acknowledgments

This work has been partially supported by the Spanish DGCyT grant
AYA-2004-08260-CO3, and the Spanish MECD FPU grant
AP2000-1389. Partial support from the Comunidad de Madrid, under 
grant S0505/ESP/000237 (ASTROCAM) is acknowledged.
AID acknowledges support from the Spanish MEC through a
sabbatical grant PR2006-0049 and thanks the hospitality of the Institute of
Astronomy of Cambridge.
RG acknowledges funding from NASA/STScI grant
HST-GO-08678.04-A and LTSA NAG5-11635. MAB also acknowledges
support from NASA/STScI grants HST GO-8678, GO-9126, GO-7675 and AR-9917, and
NSF/AST-0607516. We also thank an anonymous referee for his/her
interest in improving the readability and scientific quality of the text.

\end{document}